%% file: newformat_dimer_square_cross15.tex
\documentclass[showpacs,preprintnumbers,superscriptaddress,aps,pre,10pt]{revtex4}

\usepackage{subfigure}
\usepackage{amssymb}
\usepackage{amsmath}
\usepackage{times}
\usepackage{graphicx}
\usepackage{epsfig}

\RequirePackage{color}

\definecolor{MyDarkGreen}{rgb}{0.02,0.60,0.06}

\graphicspath{{figures}}

\graphicspath{{figures}}

\begin{document}

\newcommand{\ket}[1]{\ensuremath{|#1\rangle}}
\newcommand{\bra}[1]{{\langle #1|}}
\def\avg#1{\left\langle#1\right\rangle}
\def\bra#1{\left\langle#1\right|}
\def\ket#1{\left|#1\right\rangle}
\def\Eq#1{Eq. {\eqref{#1}}}
\def\s{\sigma}
\def\t{\tau}
\def\d{\delta}
\def\be{\begin{equation}}
\def\ee{\end{equation}}
\def\o{\omega}
\def\bea{\begin{eqnarray}}
\def\eea{\end{eqnarray}}
\def\L{{\cal L} }

\title[]{Entanglement studies of resonating valence bonds on the frustrated square lattice}

\date{November 28, 2017}

\author{Julia Wildeboer}
\email{jwildeboer@uky.edu}
\affiliation{Department of Physics \& Astronomy, University of Kentucky, Lexington, Kentucky 40506-0055, USA}

\author{Alexander Seidel}
\email{seidel@physics.wustl.edu}
\affiliation{Department of Physics, Washington University, St. Louis, Missouri 63130, USA}

\begin{abstract}
We study a short-range resonating valence bond (RVB) wave function with diagonal links on the square lattice that permits sign-problem free wave function Monte-Carlo studies. Special attention is given to entanglement properties, in particular, the study of minimum entropy states (MES) according to the method of Zhang et. al. [Physical Review B {\bf 85}, 235151 (2012)]. We provide evidence that the MES associated with the RVB wave functions can be lifted from an associated quantum dimer picture of these wave functions, where MES states are certain linear combinations of eigenstates of a 't Hooft ``magnetic loop''-type operator. From this identification, we calculate a value consistent with $\ln(2)$ for the topological entanglement entropy directly for the RVB states via wave function Monte-Carlo. This corroborates the $\mathbb{Z}_{2}$ nature of the RVB states. We furthermore define and elaborate on the concept of a ``pre-Kasteleyn'' orientation that may be useful for the study of lattices with non-planar topology in general. 
\end{abstract}


\maketitle

\section{Introduction}\label{introduction}

Frustrated quantum antiferromagnets are believed to harbor exotic states of matter known as 
quantum spin liquids. In their original guise, they were envisioned by Anderson \cite{Anderson73} to retain the full space group symmetry of some underlying lattice {\em as well as} global $SU(2)$ spin-rotational symmetry in their ground state. This gave rise to a picture where ground states are thought of in terms of fluctuating valence bond configurations, or resonating valence bonds (RVBs) (see \cite{PALEE} for a review). While originally it proved challenging to stabilize this scenario in models of local microscopic interactions, the assumption alone that this is possible has lead to deep insights regarding the role of topology in certain situations. Specifically, in the short-range RVB scenario, where all singlet bonds are limited in length (and which we will always refer to simply as ``RVB'' in the following), a four-fold ground state degeneracy has been predicted for toroidal topology, in addition to the presence of fractionalized excitations with non-trivial mutual statistics \cite{ReadChakraborty89, Kivelson89}. These are the hallmarks characterizing what was coined ``topological order'' by Wen \cite{WenTopo}, and necessitate the presence of ``long-range entanglement'' (see \cite{misguich_book} and \cite{sachdev15} for reviews of these concepts in the present context).

To construct exactly solvable models whose ground states have the same essential topological features as the RVB-wave functions, Rokhsar and Kivelson introduced so called quantum dimer models (QDMs) \cite{Kivelson89}. 
Their construction principle is based on two assumptions that simplify the setting greatly: 1. The essential physics  of quantum antiferromagnets in the targeted phase is well-captured by states containing only near (usually near{\em est}) neighbor bonds (``dimers''). 2. It is still captured when the inner product of the original spin-1/2 problem is changed so as to render certain loop dynamics Hermitian, and at the same time, render different dimer configurations orthogonal. 
Especially the latter assumption must be regarded as highly non-trivial, as any two of the original valence-bond configurations will have non-zero overlap (though they have been shown to be linearly independent on many lattices \cite{CCK, seidel09, wildeboer11}). 
Over time, however, these two assumptions have passed much scrutiny, as we will review in the following. The main purpose of this paper is to test thus far uncharted territory of this dimer-RVB correspondence. 

Irrespective of their relations with spin-1/2 RVB states, the study of ground states of QDMs at exactly solvable points has brought to light rich and interesting phase diagrams. 
At such ``Rokhsar-Kivelson'' (RK) points, the ground states of quantum dimer models tend to be equal amplitude superpositions of all dimer configurations (with possible restrictions to topological sectors). 
Counter to intuition, the original construction on the square lattice {\em failed} to stabilize a gapped liquid, but led to a critical point between many symmetry broken phases \cite{sachdev89, leung96}. Similar findings were made for a QDM on the honeycomb lattice \cite{MoessnerSondhiChandra01}.
Moessner and Sondhi correctly attributed this failure to produce a gapped topological phase to the bipartiteness of these lattices, and demonstrated that such a phase exists in a QDM on the triangular lattice \cite{MS}. The dichotomy between the phase diagrams of QDMs on bipartite and non-bipartite lattices can be understood in terms of powerful mappings between QDMs and lattice gauge theories \cite{MoessnerSondhiFradkin, Fradkin_book} (see also \cite{Ng} for a recent review), which are exact in some cases \cite{misguich}. 

A non-trivial question now arises when considering the RVB-counterparts of RK-type wave functions, that is, equal amplitude superpositions of all nearest neighbor valence bond coverings in some lattice topology. One may ask what properties of RVB RK-states 
are faithfully captured by the well-understood quantum dimer RK-states. For the latter, correlation functions are often analytically known in terms of Pfaffians \cite{wangwu}. On the other hand, for RVB RK-states on some {\em bipartite} lattices, pertinent studies have a long tradition \cite{doucout, alet, tang}, following a loop-gas Monte-Carlo method due to Sutherland \cite{sutherland}. These studies have confirmed the critical behavior of RVB RK-states respecting a bipartite structure \cite{alet, tang} first inferred from QDM RK-states, albeit with somewhat different critical exponents. In the non-bipartite case, defining physical properties of RVB RK-states have only been calculated more recently, owing to a sign-problem of the loop-gas method in this case. 
\begin{figure}[tbp]
\includegraphics[width=0.28\textwidth]{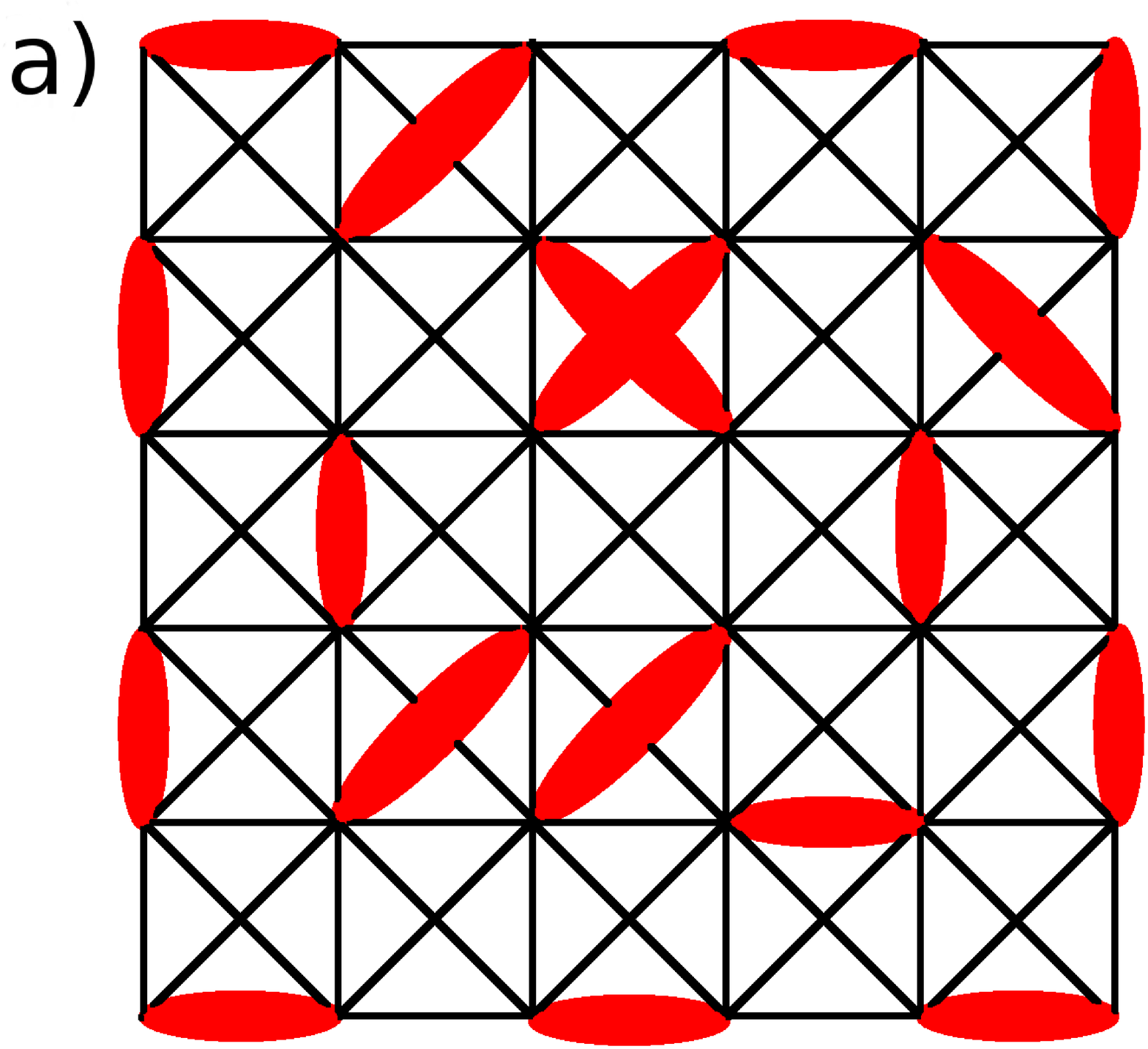}\quad \quad \quad \quad
\includegraphics[width=0.28\textwidth]{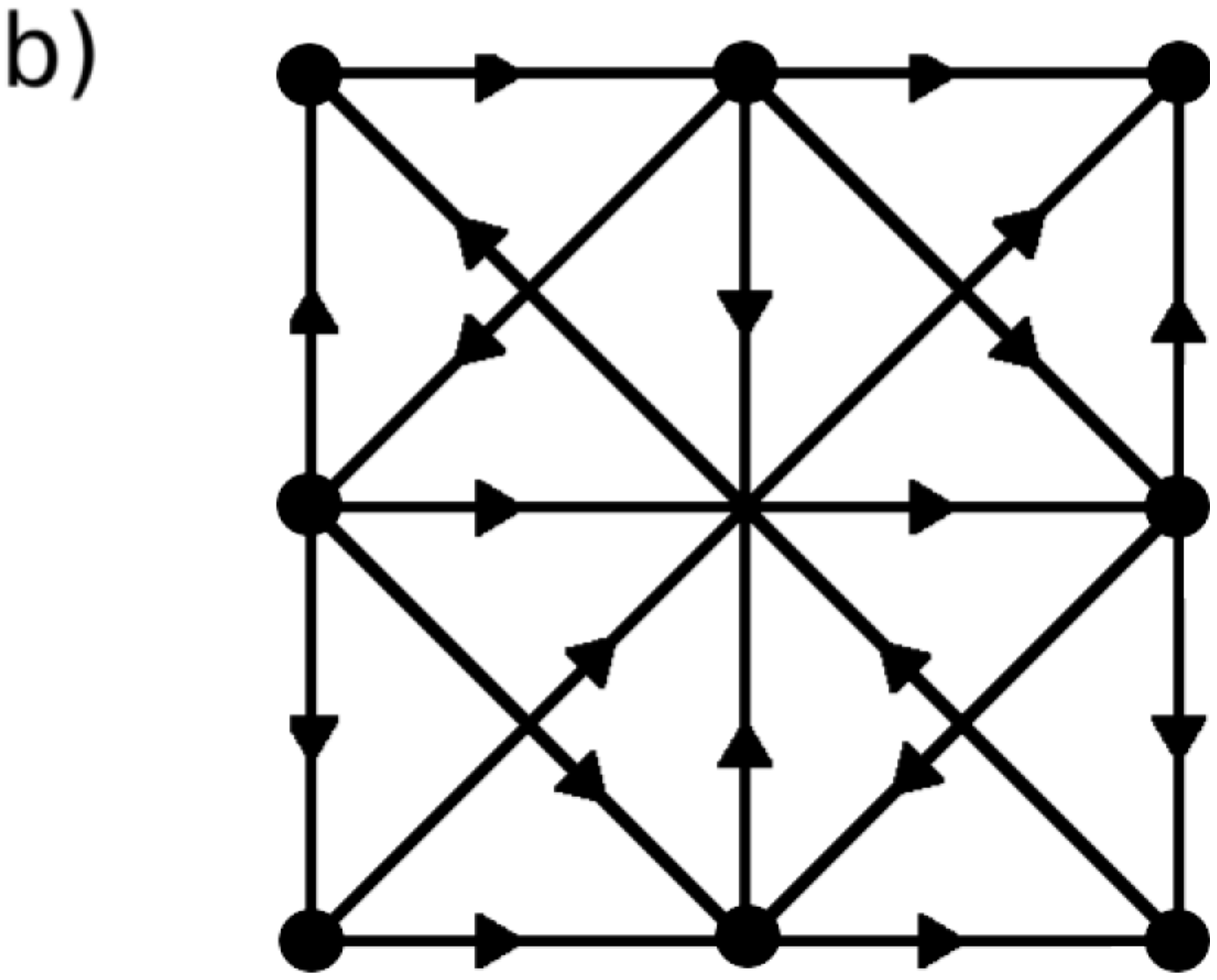}\quad \quad \quad \quad
\includegraphics[width=0.27\textwidth]{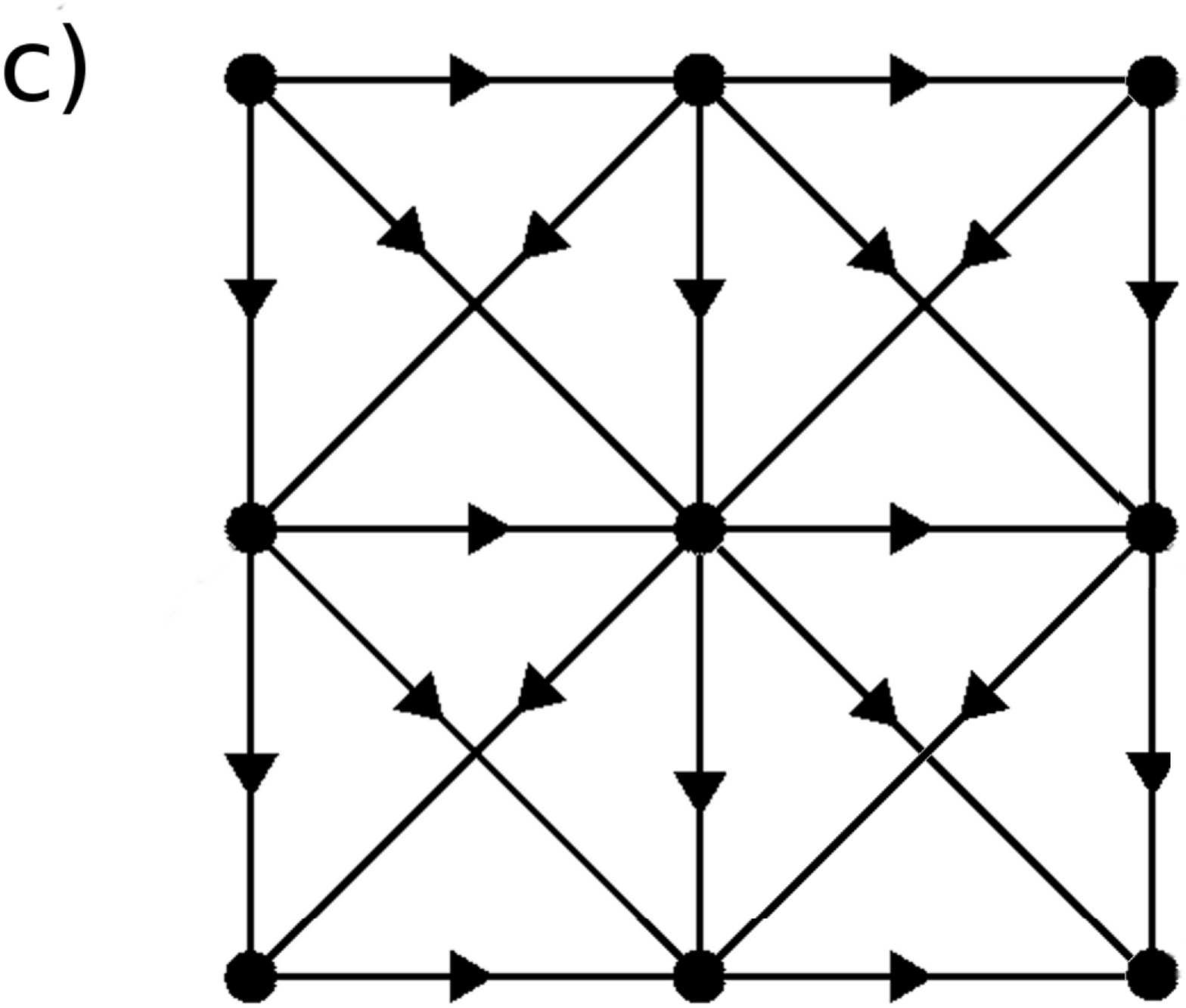}
\caption{
a) A dimer covering on the frustrated square lattice. Note that one plaquette is internally dimerized with a pair of crossed valence bonds.
This covering appears in \eqref{state} with a negative amplitude. 
b) The ``pre-Kastelyn'' orientation of the lattice discussed in the main text. This orientation 
is used to in the Pfaffian re-formulation of the wave function \eqref{state}. Note that this artificially doubles the unit cell, but this is without physical consequence (see c)).
c) The link orientation used to define the sign of valence bonds in \Eq{state}. Note that this orientation 
manifestly preserves the full translational symmetry of the square lattice, thus ensuring that the wave function \Eq{state} has this symmetry. 
}
\label{figure_1}
\end{figure}
In Ref.\cite{wildeboer12}, we developed a Pfaffian presentation of some RVB RK-states on the triangular and kagome lattice, which allows sign-problem free Monte Carlo studies. This method was later used to calculate entanglement properties, and also constrained the modular $\mathcal{S}$ and $\mathcal{U}$ matrices \cite{wildeboer15}.
In all, these studied have strongly corroborated to hypothesis that these wave functions are in the same $\mathbb{Z}_2$ topological phase as their QDM counterparts \cite{MS, misguich}. 

Historically, the square lattice has played an important role in motivating the search for spin liquid physics, owing to its relative experimental abundance, notably in the high-T$_c$ parent compounds.
Indeed, the frustrated geometries of non-bipartite lattices, which have produced a number of exciting experimental candidates for quantum spin liquids \cite{Shimizu, Itou, Helton, Okamoto, Feng}, are not expected to be necessary to access these physics, as interactions beyond nearest neighbors can introduce frustration in any geometry. In this context, an interesting prediction of QDMs is that a topological $\mathbb{Z}_2$ liquid phase {\em can} be accessed at the critical RK-point of the square lattice QDM as soon as non-zero amplitudes are introduced \cite{paul} for next-nearest neighbor dimers, residing on the diagonals of the lattice (Fig. \ref{figure_1}a).  {\em The primary purpose of this paper is to consider 
RVB-wave functions with diagonal bonds 
on the square lattice, which admit a Pfaffian presentation and are thus amenable to Monte-Carlo studies of its correlations \underline{and} entanglement properties.} We demonstrate that such wave functions indeed lie in the expected $\mathbb{Z}_2$ topological phase. 

In closing this section, we briefly remark on the status of parent Hamiltonians for the $SU(2)$-invariant RVB RK-states. Indeed, these states do not easily lend themselves to the construction of local parent 
Hamiltonians, fueling controversy for some time. 
The earliest successful stabilizations of  $\mathbb{Z}_2$ topological phases have indeed done away with $SU(2)$-invariance, as evidenced by the non-bipartite QDMs, as well as the celebrated Kitaev toric code \cite{toric}, all of which lie in the phase described by deconfined (weak coupling) Ising gauge theory in three dimensions \cite{wegner}. By now, however, various costruction principles for RVB RK-states have been discussed. 
In the limit of highly decorated lattices, the construction of RVB-stabilizing Hamiltonians has been argued to be perturbatively under control \cite{RamanMoessnerSondhi}. In the following, we will focus on {\em simple} lattices. 
For the bipartite case, (critical) RVB RK-states have first been stabilized by Fujimoto \cite{fujimoto}, with some simplifications 
(and, in one case, corrections) later given by Cano and Fendley \cite{cano}. The kagome (topological) RVB RK-state referenced above is stabilized by the parent Hamiltonian constructed by one of us \cite{seidel09}. Subsequent work has, moreover, proven the uniqueness of the ground states of this Hamiltonian, modulo four-fold topological degeneracy on the torus \cite{schuch, zhou}. 
The approach of Ref. \cite{cano} may also be applied to certain non-bipartite lattices, where, however, for the ground state uniqueness one relies (thus far) on the ability of Klein models \cite{klein} to gap out (at least on finite lattices) non-nearest-neighbor valence bond states. This property of Klein models appears to be well established only for some bipartite lattices \cite{CCK}, does not hold for the kagome or triangular lattice, but is at least {\em expected} to hold for sufficient lattice decoration \cite{RamanMoessnerSondhi}.

The remainder of this paper is organized as follows: 
In Sec. \ref{RVBdef}, we briefly introduce the RVB wave function that will be investigated in this paper.
in Sec. \ref{MCmethod}, we review the Pfaffian Monte-Carlo method of Refs. \cite{wildeboer12, wildeboer15} in some detail, and set the stage for its application to a situation with a non-planar lattice graph. To this end, the notion of a ``pre-Kasteleyn'' orientation is introduced, and various statements are proven. We then present the results of the application of these methods to the calculation of correlation functions and Renyi entropy, confirming the wave function's low energy effective theory to be the $\mathbb{Z}_2$ topological phase. In Sec. \ref{QDMSEC} we define a quantum dimer model related to our RVB wave function. We summarize and conclude in Sec. \ref{conclusion}. An appendix elaborates on some defining properties of Kasteleyn-like orientations.

\section{The wave function for the nonbipartite square lattice \label{RVBdef}}

In this work, we aim to analyze the correlation functions and the Renyi entanglement entropy of an RVB state defined on a square lattice topology including diagonal links, as shown in Fig. \ref{figure_1}a). We will refer to this topology as a nonbipartite and/or frustrated square lattice. 
The RVB wave function to be considered is defined via 
\begin{eqnarray}
|{\sf RVB} \rangle = \sum_{D}(-1)^{n_{c}(D)}|D\rangle\;.
 \label{state}
\end{eqnarray}
Here, $D$ goes over all possible ``dimerizations'' or dimer coverings of the lattice, i.e., pairings of the 
lattice into neighboring pairs along the links of the non-bipartite square topology. 
A typical dimer covering is shown in Figure \ref{figure_1}(a). 
Each lattice site is equipped with a spin-$1/2$ degree of freedom. For each dimer covering 
$D$, $|D\rangle$ denotes a state where each dimer of the covering $D$ is realized by a singlet between the associated spins, 
and a sign convention is used that corresponds to an orientation of links 
as shown in  Figure \ref{figure_1}(c). 
We point out that our lattice topology does not correspond to a planar graph. 
This precludes using methods of Kasteleyn \cite{kasteleyn} in their original form, which were 
quite essential in our Pfaffian presentation of RVB wave functions on the kagome and triangular lattices \cite{wildeboer12}. 
However, we find that a generalization of these methods is possible if we furnish the 
definition of the RVB-state with certain additional phases as shown in \Eq{state}. Here, 
the parameter $n_{c}(D)$ in \eqref{state} counts the number of crossings between two dimers 
that a particular dimer covering contains. 
Thus, the dimer coverings with an even number of crossed pair of dimers, $n_{c} = 0, 2, \ldots$ are equipped with a positive amplitude,  
whereas an odd number of crossing, $n_{c} = 1, 3, \ldots$, in a dimer covering leads to a negative amplitude. 
We emphasize that despite the need to introduce additional phases owing to the departure from planar graph topology, the rule that determines these phases is still {\em local}, and can, as we will show, in particular be obtained from a quantum dimer model with local interactions. We may therefore regard \Eq{state} as a natural variation of the RVB RK-state.
We will now proceed to show in detail how the Pfaffian Monte-Carlo scheme of Refs. \cite{wildeboer12, wildeboer15} can successfully be applied to 
this wave function \Eq{state}.

\section{Pfaffian Monte Carlo and Correlations and Entanglement Entropy\label{MCmethod}}
In order to make this work self-contained, we will now briefly review the Pfaffian Monte Carlo technique. 
The standard method for calculating expectation values, e.g., for the product of two local operators
${\cal O}_1 {\cal O}_2$, 
for {\em bipartite} 
lattices is due to Sutherland \cite{sutherland}.
It is based on the observation matrix elements of the form 
$\langle {\sf RVB}|{\cal O}_1 {\cal O}_2|{\sf RVB}\rangle$ will, quite generally, only depend on the configuration of close-packed 
non-intersecting loops associated to the overlap graph between $D$ and $D'$ (that is, basically, the union of the sets of all dimers 
in $D$ and $D'$, which can always be regarded as the disjoint union of close-packed non-intersecting loops). 
This maps the correlator 
\begin{eqnarray}\label{corr}
\cfrac{\langle {\sf RVB}|{\cal O}_1 {\cal O}_2|{\sf RVB}\rangle}{\langle {\sf RVB}|{\sf RVB}\rangle}
=\cfrac{\sum_{D,D'}\langle D|{\cal O}_1 {\cal O}_2|D'\rangle}{\sum_{D,D'}\langle D|D'\rangle}\,.
\end{eqnarray}
onto a {\em classical} loop gas problem, provided that all inner products $\langle D|D'\rangle$ are non-negative. In contrast, 
this is in general not the case for non-bipartite lattices: 
Typically, no sign convention for the states $\ket{D}$ exists that has this property. (Note that it is  generally sufficient to identify three dimer covering on the lattice of interest for which the no such convention exists.) Any attempt to solve the problem by Monte-Carlo evaluation of \Eq{corr} 
will then be plagued by the sign problem.

The only way to circumvent such a sign problem  is  abandon the description in term of the $\ket{D}$-states and instead move to a different expansion of the RVB-state in terms of
states with non-negative mutual overlaps, while, {\em at the same time}, the coefficients in this expansion can be efficiently calculated. In particular, moving to any orthogonal basis precludes the appearance of negative overlaps, and moves all complexity to the problem of calculating coefficients.
This is the route explored in Refs.\cite{wildeboer12,wildeboer15} for kagome and triangular lattice RVB RK-states, and will be established here for the ``non-planar'' square lattice RVB \Eq{state}.
To this end, we consider the Ising-basis of local $S_z$-eigenstates $\ket{I}$ and write $\ket{{\sf RVB}}= \sum_I a_I \ket{I}$, the sum running over all Ising configurations. We may then write
\begin{eqnarray}
  \label{corr2}
\frac{\langle {\sf RVB}| {\cal O}_i {\cal O}_j \ket{\sf RVB}}{\bra{\sf RVB}{\sf RVB} \rangle}  
&=& 
\frac{\sum_I \sum_{I'} a_{I} a_{I'} \bra{I'} {\cal O}_i{\cal O}_j \ket{I}}{\sum_{I} |a_{I}|^2} \nonumber \\ 
&=& 
\frac{\sum_{I} |a_{I}|^2 \sum_{I'} \frac{a_{I'}}{a_{I}} \bra{I'} {\cal O}_i {\cal O}_j \ket{I} }{\sum_{I} |a_{I}|^2}\,.  \nonumber \\
& &
\end{eqnarray}
This may now be interpreted as the classical expectation value 
of a quantity $f$:
\begin{equation}
\langle f \rangle = {\sum_{I} f_{I} e^{-E_I}}/{\sum_{I} e^{-E_I}}\;,
\end{equation}
where $e^{-E_I} = |a_I|^2$ and the value $f_I$ of the quantity $f$ in the Ising configuration $I$ is given by 
$f_{I} = \sum_{I'} \bra{I'}{\cal O}_i {\cal O}_j \ket{I} \frac{a_{I'}}{a_I}$.
It turns out that for most local operators, the sum over $I'$ only contains a few non-zero terms.
The problem is thus properly set up for Monte-Carlo evaluation, as long as the coefficients $a_I$ can be calculated. It turns out that for the choice of phases made in \Eq{state} this is the case.

For reference purposes, it is a good idea to to first consider the ``plain vanilla'' RVB state, which, similar to the original Rokhsar-Kivelson quantum dimer states, has all amplitudes set equal to one:
\begin{eqnarray}
|{\sf RK} \rangle = \sum_{D}|D\rangle =\sum_I b_I\ket{I}\;.
 \label{RK}
\end{eqnarray}
Note that for any lattice topology {\em without} crossings, there is no difference between
Eqs. \eqref{RK} and \eqref{state}. 
For any oriented lattice graph, the Ising-coefficients $b_I$ of $\ket{\sf RK}$ 
 can naturally be written as a ``Hafnian'': 
\begin{eqnarray}\label{bI}
b_{I} = \mbox{Haf}[M_{ij}(I)] \equiv\frac{1}{2^{N/2}(\frac{N}{2}!)}
\sum_{\lambda \in S_N} M_{\lambda_1 \lambda_2}(I) M_{\lambda_3 \lambda_4}(I) \times \cdots \times 
M_{\lambda_{N-1} \lambda_N}(I) \,.
\end{eqnarray}
Here, $M$ is a symmetric matrix whose indices run over the $N$ lattice sites and which depends on the Ising 
configuration via $M_{ij}(I)= \Theta_{ij}( \delta_{\sigma_{i},\uparrow} \delta_{\sigma_{j},\downarrow} - 
\delta_{\sigma_{i},\downarrow} \delta_{\sigma_{j},\uparrow})$, where the $\sigma_{i}$ 
describe the Ising configuration $I$,
$\Theta_{ij}$ encodes the orientation of links connecting nearest neighbors (NN) in the 
lattice topology via
\be
  \Theta_{ij}=
  \begin{cases} 
      1 & i<j  \\
      -1  & j<i \\
      0 & i,j\;\; \mbox{not NN} 
   \end{cases}
\ee
and ``$i<j$'', by abuse of notation, refers to a nearest neighbor pair $(i,j)$ ordered in accordance with the orientation. 
For the frustrated square lattice, we will consider the orientation 
indicated by the arrows in Figure \ref{figure_1}(c).
$\lambda$ runs over all permutations 
of the $N$ sites. 
\begin{figure}[tbp]
\includegraphics[width=0.16\textwidth]{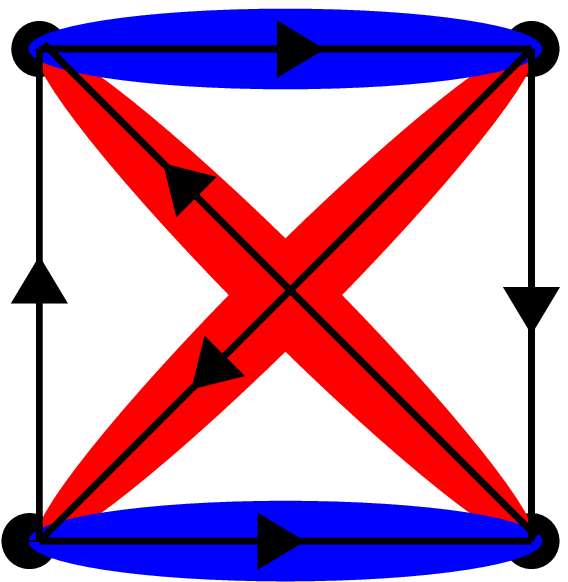}\quad \quad \quad \quad
\includegraphics[width=0.16\textwidth]{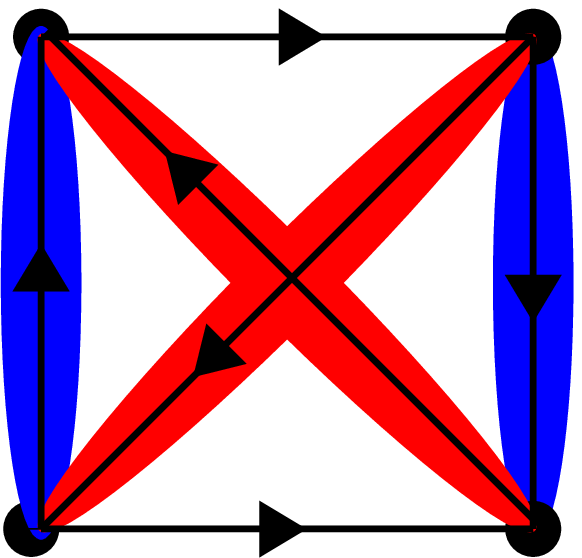}\quad \quad \quad \quad
\includegraphics[width=0.16\textwidth]{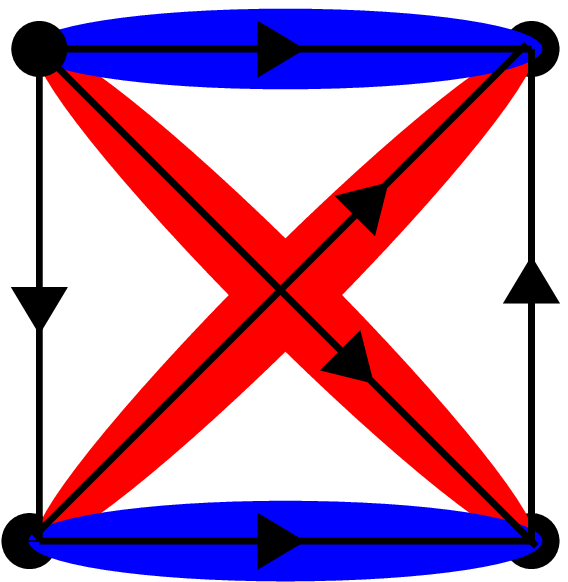}\quad \quad \quad \quad
\includegraphics[width=0.16\textwidth]{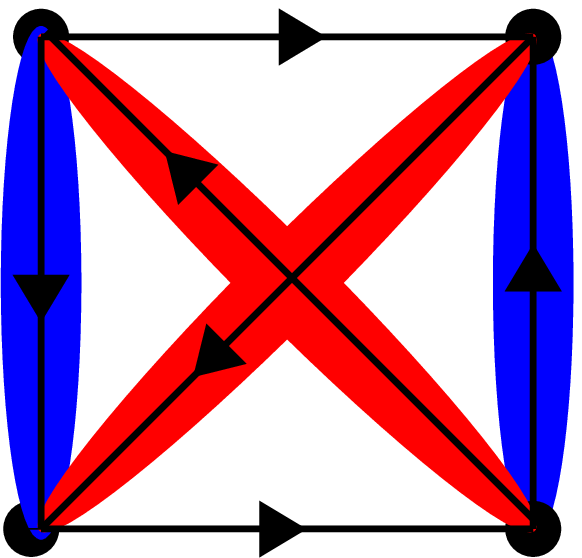}
\caption{
The possible configurations for bowtie loops on the frustrated square lattice. 
Four configurations arise because of the doubling of the unit cell by the pre-Kasteleyn orientation (shown). One observes that traversal of any of these loops in any direction always leads to an even number of arrows parallel or anti-parallel to the traversal. 
}
\label{bowtie}
\end{figure}
The correspondence between \Eq{RK} and \Eq{bI} is easy to see once one realizes that any 
$\lambda$ giving a non-zero contribution to \Eq{bI} naturally defines a dimerizarion $D$, as all 
$(\lambda_{2n-1},\lambda_{2n})$ must be nearest neighbor pairs, and the product of matrix elements in \Eq{bI} 
is then just the contribution of $\ket{D}$ to the Ising configuration $I$. The overall combinatorial factor 
in the equation compensates for a many-to-one correspondence between permutations and dimerizations.

Note that the formal definition of the Hafnian is related to that of the 
Pfaffian 
through the absence/presence of the sign factor $(-1)^\lambda$, in a manner similar to how the permanent is related to the determinant. 
While the determinant or Pfaffian can be computed efficiently in polynomial time, it is not known how to do this for the permanent or Hafnian. 
\Eq{bI} therefore does not lend itself to a Monte-Carlo scheme as discussed above. 
To make progress, let's for the time being {\em define} an RVB-state of the form 
 \be
 \begin{split}\label{RVBp}
    \ket{\sf RVB}'&= \sum_I a_I' \ket{I}\\
    a_I'&= \mbox{Pfaff} [M_{ij}(I)\tilde\Theta_{ij}]=\frac{1}{2^{N/2}(\frac{N}{2}!)}\sum_{\lambda \in S_N} 
  (-1)^\lambda  \,\tilde \Theta _{\lambda_1 \lambda_2} \times\dotsc\times \tilde \Theta _{\lambda_{N-1} \lambda_N}\,
    M_{\lambda_1 \lambda_2}(I) \times \cdots \times 
M_{\lambda_{N-1} \lambda_N}(I)
 \end{split}
 \ee
Here, $\tilde \Theta_{ij}$ is an orientation matrix describing yet another orientation of the lattice (usually different from $\Theta_{ij}$).  
Ordinarily, one requires $\tilde \Theta_{ij}$ to be a so-called 
 ``Kasteleyn orientation'', which always exists for a planar lattice graph \cite{kasteleyn}. 
 Presently, however, we will also need to admit somewhat more general orientations $\tilde \Theta_{ij}$, e.g., 
for the frustrated square lattice as shown in Fig. \ref{figure_1}b). 
The state \eqref{RVBp} is related to the $\ket{\sf RK}$-state in \Eq{RK} via the additional phases 
$q(\lambda)\equiv q(D)= (-1)^\lambda  \,\tilde \Theta _{\lambda_1 \lambda_2} \times\dotsc\times \tilde \Theta _{\lambda_{N-1} \lambda_N}$, 
which, as on may easily see, only depend on the dimerization $D$ associated to $\lambda$. To better understand those phase, one may note that $q(D)$ and $q(D')$ are simply related whenever $D$ and $D'$ are related to each other by a single ``resonance move'' along a contractable closed loop of even length. That is, there exists such a  loop, whose links alternatingly belong to $D$ and $D'$, respectively, while all dimers not along the loop are the same for $D$ and $D'$. Our definition of a loop also entails that each link may be traversed only once. 
\begin{figure}[tbp]
\includegraphics[width=0.3\textwidth]{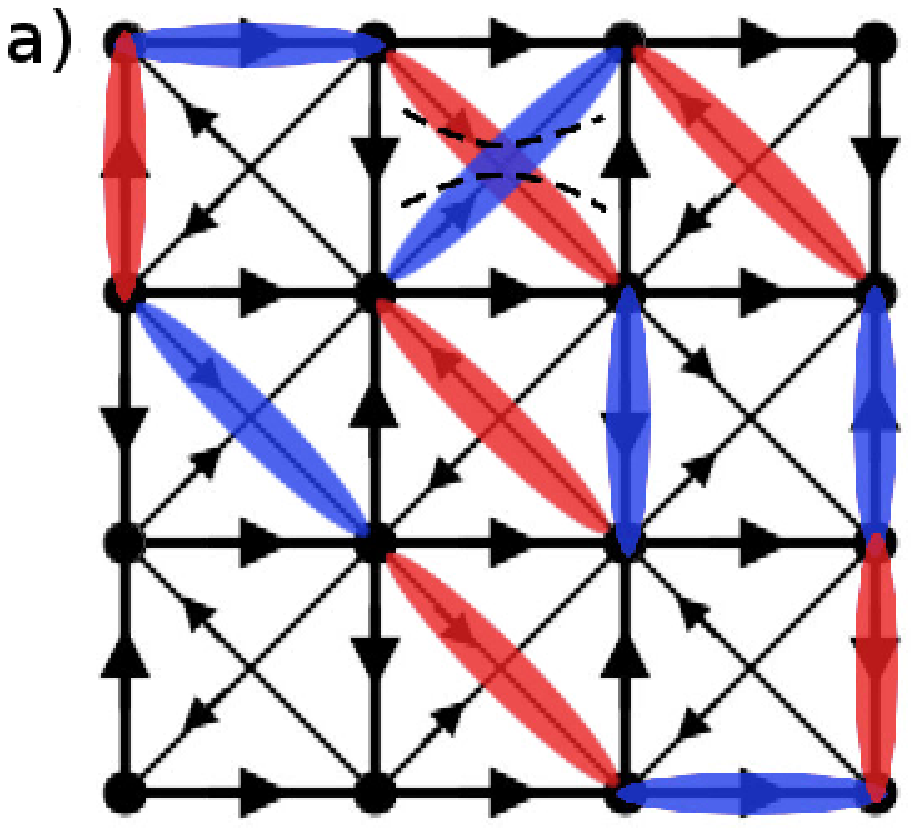}\quad \quad \quad \quad \quad \quad \quad \quad \quad
\includegraphics[width=0.3\textwidth]{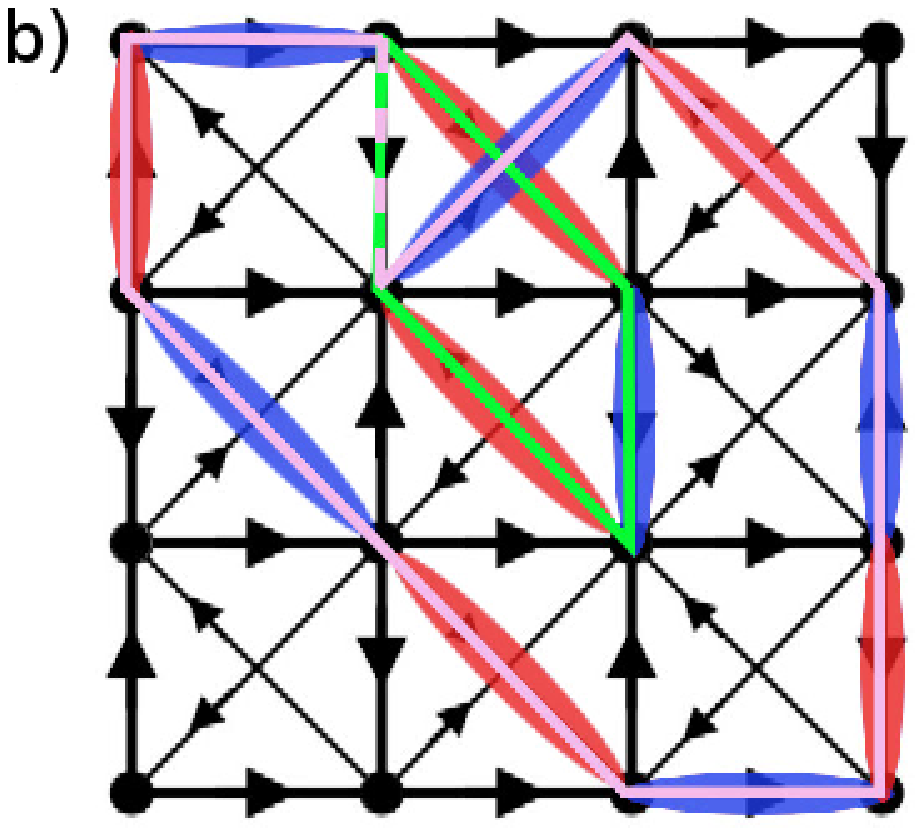}
\caption{
Illustration of Lemma 3. The orientation $\tilde\Theta$ entering the Pfaffian formulation is shown.
Only those dimers are shown that belong to a certain self-intersecting loop in the overlap of coverings $D$ (red) and $D'$ (blue). These dimers form a self-intersecting loop with both $D$ and $D'$ contributing to the cross.
a) A cut with subsequent re-gluing is indicated (dashed line) that splits the single loop into two closed loops with the crossing eliminated. These two loops do not yet correspond to legitimate resonance moves, as the diagonal links of the cross are ``broken'' apart. b) A deformation of the two separate loops defined in a) that leads to valid resonance moves, which have the same effect as the single self-intersecting resonance move defined by the original dimer loop. Loops are indicated by pink and green lines, respectively. Note that these loops share one link (dashed, west of the cross), which is unoccupied for both $D$ and $D'$. When transforming $D$ into $D'$ (red into blue), the smaller, length 4 resonance move must be carried out first (green), followed by the other move (pink).
}
\label{selfcross}
\end{figure}
On any finite lattice, it makes sense to define a ``topological sector'' of dimerizations as an (equivalence) class of dimerizations related 
to each other by sequences of resonance moves along arbitrary contractible loops. We will then find that all phases $q(D)$ are simply related 
for dimerizations belonging to the same topological sector. The key to these results is the following 
property, somewhat weaker but more general than that of a true Kasteleyn orientation:

{\underline{Definition:} \em 
We refer to a two-dimensional lattice endowed with an oriented link topology as ``pre-Kasteleyn''
if it has the following property: For any closed, non-intersecting, contractible loop along edges (links), the parity of 
the number of 
clockwise oriented edges is opposite to the parity
of the number of sites enclosed by the loop. 
} 

Note that being pre-Kasteleyn is not an intrinsically ``graph theoretical'' property, since notions such as the ``enclosed sites'' will depend on the particular embedding of the graph into the plane or the torus\footnote{Even for a planar graph, these notions become meaningful only once a designated ``outer face'', or boundary, is chosen.}. (So do the notions of ``crossing'' links and ``contractible''.) That is why 
it is preferable in this context to think of the structure in question as a ``lattice with a topology'' rather than an abstract graph, which is in any case natural in the present physical situation: The term lattice will always imply that there is meaning to the positions of the lattice sites and links on some two-dimensional manifold (plane or torus). In this spirit, we will speak of a ``lattice with a planar topology'' if after the embedding of the lattice graph into the two-dimensional manifold, there are no crossings between different links. In particular, this does \underline{not} preclude toroidal topology.


If, now, the lattice is equipped with a planar topology, the notion of pre-Kasteleyn defined above becomes equivalent to Kasteleyn, which we define as usual via

{\underline{Definition:} \em 
A lattice with an oriented planar topology is ``Kasteleyn'' if the number of clockwise edges around each face is odd.} 

We establish the equivalence of these notions for planar lattice graphs in
 Appendix \ref{app}. The notion of ``pre-Kasteleyn'' as defined above is thus a very natural
 generalization of Kasteleyn-oriented lattices in the planar case.
 For all our purposes, the essential difference between pre-Kasteleyn and ``true'' Kasteleyn lies only in the fact that, on any planar lattice graph, any loop that can define a legitimate resonance move for any dimerization of the lattice necessarily  encloses an even number of sites (all the enclosed sites form dimers with one another). This is not so in the general pre-Kasteleyn situation.
 
  The key observation for the remainder of this section is that the frustrated square lattice with the orientation $\tilde \Theta$ of Fig. \ref{figure_1}b) is pre-Kasteleyn. This, too, is proven in the Appendix, where we also discuss a simple general criterion.  
 The following discussion will focus on the frustrated square lattice with the orientation of Fig. \ref{figure_1}b).
Results simplify in obvious ways for the usual Kasteleyn-orientated planar lattice graphs, and are also expected to generalize readily to other situations with pre-Kasteleyn orientations. Obvious generalizations include the checkerboard lattice graph obtained by omission of diagonal links on every second plaquette.
For the problem at hand, we now intend to show that for $D$, $D'$ in the same topological sector, 
 \be \label{qphase}
   \frac{q(D)}{q(D')} = (-1)^{n_c(D)-n_c(D')}\;.
 \ee
This is equivalent to saying that \Eq{RVBp} equals \Eq{state} up to a sign that may depend on the topological sector. We will not care about relative phases between different topological sectors, as local operators would be oblivious to such phases \cite{misguich}.

We demonstrate \Eq{qphase} via three small lemmas: 

{\em Lemma 1:} \Eq{qphase} holds if $D$ and $D'$ are related by a single resonance move along a loop that is not self-intersecting. 

 Proof: For a single loop, we may choose $\lambda$ and $\lambda'$ associated to $D$, $D'$ such that 
 $\lambda'=\mu \circ\lambda$, where $\mu$ is a cyclic permutation of lattice sites along the loop. 
 Then it is easy to see that $ \frac{q(D)}{q(D')}=- (-1)^{N_\circlearrowright}$, where $N_\circlearrowright$ is the number of clockwise arrows encountered for the $\tilde \Theta$ orientation along the loop (or counter-clockwise, note that resonance loops are necessarily even in lengh). By the defining property of 
 pre-Kasteleyn lattice graphs (cf. Appendix \ref{app}), $N_\circlearrowright=1+N_{\sf in} \mod 2$, where $N_{\sf in}$ is the number of sites enclosed by the loop. 
Moreover, $N_{\sf in}=N_{\sf x} \mod 2$, where $N_{\sf x}$ is the number of dimers in both $D$ and $D'$ that are crossing the loop. (Any such dimer is not part of the loop, which is free of self-intersections. Therefore, since no link of the frustrated square lattice crosses more than one other link, any such dimer has precisely one of its sites enclosed by the loop.)
 Finally, $N_{\sf x}=N_{\sf x}(D)+N_{\sf x}(D')$, where $N_{\sf x}(D)$ and $N_{\sf x}(D')$ are the crossed dimers {\em belonging to the loop} associated with $D$ and $D'$, respectively. Modulo 2, this is the same as $N_{\sf x}(D)-N_{\sf x}(D')=n_c(D)-n_c(D')$. This proves Lemma 1. 

\begin{figure}[tbp]
\includegraphics[width=0.666\textwidth]{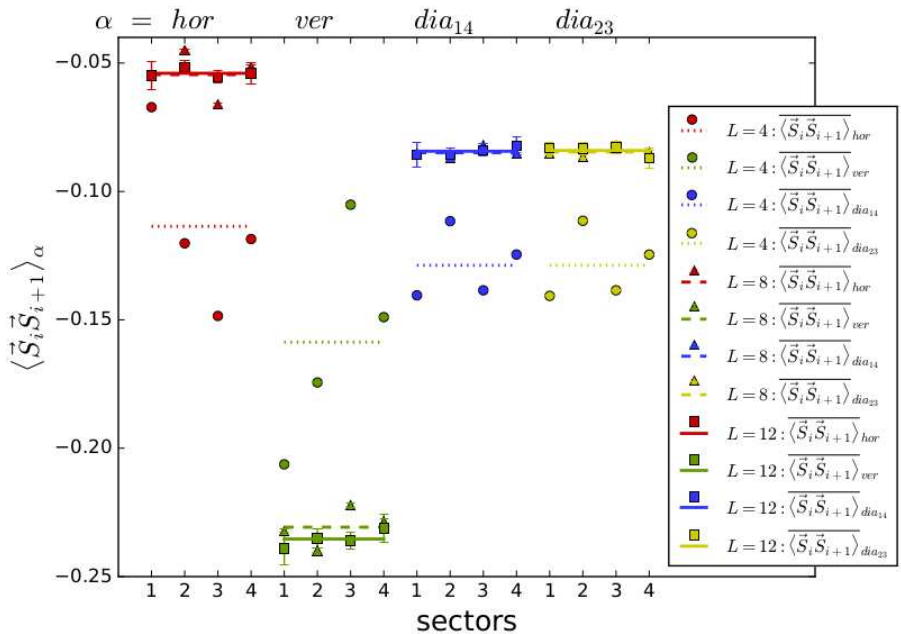}
\caption{Expectation values for various local singlet operators for different lattice sizes $L\times L$. 
Shows are values for each of the projections of the wave function \eqref{state} onto the four topological sectors. For the smallest, $L=4$ lattice (circles), strong discrepancies between sectors are visible for each of the local bond operators (horizontal, vertical, and the two types of diagonal bonds.) These discrepancies decrease rapidly with system size (triangles and squares for $L=8,12$, respectively). 
Dotted, dashed, and solid lines indicate the averages for the different sectors for the different respective lattice sizes. Note that the link orientation of Fig. \ref{figure_1}c) does not preserve $\pi/2$ degree rotational symmetry. This is clearly shown by the difference between vertical and horizontal bonds. Remaining mirror symmetries relate the different diagonal links, however. This, too, is clearly seen at all system sizes. }
\label{average_heisenberg}
\end{figure}
\begin{table}[t]
    \begin{tabular}{ | c | c || r | r | r | r |}
    \hline
  $L$ & $sec$& $\langle \vec{S}_{i} \vec{S}_{i+1}\rangle_{hor}$ & $\langle \vec{S}_{i} \vec{S}_{i+1}\rangle_{ver}$ & $\langle \vec{S}_{i} \vec{S}_{i+1}\rangle_{dia_{14}}$ & $\langle \vec{S}_{i} \vec{S}_{i+1}\rangle_{dia_{23}}$          \\ \hline \hline
    4 & $(e,e)$ & -0.06718 $\pm$ 0.00019     & -0.20628 $\pm$ 0.00017     & -0.14045 $\pm$ 0.00015       & -0.14066 $\pm$ 0.00015     \\ \hline
    4 & $(o,e)$ & -0.12020 $\pm$ 0.00011     & -0.17439 $\pm$ 0.00011     & -0.11156 $\pm$ 0.00009       & -0.11142 $\pm$ 0.00013     \\ \hline
    4 & $(e,o)$ & -0.14849 $\pm$ 0.00015     & -0.10516 $\pm$ 0.00011     & -0.13851 $\pm$ 0.00011       & -0.13849 $\pm$ 0.00016     \\ \hline
    4 & $(o,o)$ & -0.11855 $\pm$ 0.00013     & -0.14896 $\pm$ 0.00012     & -0.12456 $\pm$ 0.00012       & -0.12459 $\pm$ 0.00015     \\ \hline \hline
    8 & $(e,e)$ & -0.05506 $\pm$ 0.00118     & -0.23239 $\pm$ 0.00103     & -0.08543 $\pm$ 0.00087       & -0.08517 $\pm$ 0.00066     \\ \hline
    8 & $(o,e)$ & -0.04509 $\pm$ 0.00071     & -0.23987 $\pm$ 0.00082     & -0.08701 $\pm$ 0.00056       & -0.08645 $\pm$ 0.00062     \\ \hline
    8 & $(e,o)$ & -0.06617 $\pm$ 0.00064     & -0.22224 $\pm$ 0.00072     & -0.08202 $\pm$ 0.00055       & -0.08219 $\pm$ 0.00057     \\ \hline
    8 & $(o,o)$ & -0.05168 $\pm$ 0.00088     & -0.22849 $\pm$ 0.00119     & -0.08523 $\pm$ 0.00056       & -0.08471 $\pm$ 0.00062     \\ \hline \hline
   12 & $(e,e)$ & -0.05482 $\pm$ 0.00541     & -0.23911 $\pm$ 0.00633     & -0.08569 $\pm$ 0.00481       & -0.08308 $\pm$ 0.00224     \\ \hline
   12 & $(o,e)$ & -0.05151 $\pm$ 0.00259     & -0.23516 $\pm$ 0.00359     & -0.08559 $\pm$ 0.00234       & -0.08334 $\pm$ 0.00233     \\ \hline
   12 & $(e,o)$ & -0.05547 $\pm$ 0.00245     & -0.23593 $\pm$ 0.00323     & -0.08387 $\pm$ 0.00234       & -0.08284 $\pm$ 0.00251     \\ \hline
   12 & $(o,o)$ & -0.05405 $\pm$ 0.00401     & -0.23120 $\pm$ 0.00531     & -0.08221 $\pm$ 0.00352       & -0.08689 $\pm$ 0.00394     \\ \hline
    \end{tabular}
\caption{Raw data of local operator expectation values in Fig. \ref{average_heisenberg}. It is seen that these local operators cannot distinguish different topological sectors at sufficient system sizes.
}
\label{tab:table0}
\end{table}
{\em Lemma 2:} \Eq{qphase} holds if $D$ and $D'$ are related by a single ``bowtie'' resonance move, i.e., the shortest (length 4) self-intersecting resonance move taking two parallel dimers into a cross and vice versa (Fig. \ref{bowtie}). 

Proof: Following the logic of the proof of Lemma 1, it is now to be shown that if we traverse the bowtie  in either direction, we encounter an even number of arrows that are oriented opposite to the direction of the traversal. One may easily check directly that the $\tilde \Theta$-orientation has this property, as shown in Fig. \ref{bowtie}). 

{\em Lemma 3:} Any two dimer coverings $D$ and $D'$ that are each without crosses can be related by a sequence of resonance moves along non-self-intersecting loops. 

Proof: Note that the loops forming the overlap graph between $D$ and $D'$ may still intersect, and each individual loop may be self-intersecting. However, unlike in the situation of lemma 2, at each cross exactly one dimer belongs to $D$ while the other belongs to $D'$. The  situation is illustrated in Fig. \ref{selfcross}. If we follow the branch emerging from the South-West corner of the crossing, the next corner of the crossing we arrive at cannot be North-East. For otherwise, the crossing would be between two separate loops. The figure assumes (without loss of generality) that starting South-West, we first arrive again at the South-East corner. Then, a horizontal cut and re-gluing as shown in  Fig. \ref{selfcross}a) will split the loop into two separate loops, which are, of course, not corresponding to legitimate resonance moves yet (a corner occurs at the center of a plaquette). However, the two separate loops in Fig. \ref{selfcross}b) correspond to two legitimate resonance moves, which, when carried out in the right order, reproduce the effect of a resonance move along the original single loop (see caption). The two new loops have in total one less self-intersection than the original loop. One may thus proceed inductively. 

Together, these three lemmas prove \Eq{qphase}, as one can of course connect any two dimer patterns $D$ and $D'$ first to ones without crosses via bowtie moves, and can then connect the resulting patterns via non-self-intersecting resonance moves because of Lemma 3. By Lemmas 1 and 2, \Eq{qphase} applies to each step of this procedure, and so applies to $D$ and $D'$ as well. 

Note that in the standard situation of planar graphs and Kasteleyn orientations, Lemma 1 is all one needs, and moreover, absent any momomers, 
$N_{\sf in}$ is automatically even. 

In the following, we will ignore the topological sector dependent phases mentioned above, and drop the primes in \Eq{RVBp}. 
We then have the possibility to compute the coefficients $a_I$ in polynomial time, and feed them into a sign-problem-free Monte-Carlo scheme via \Eq{corr2}. That the frustrated square lattice RVB wave function \Eq{state} indeed has this property was already pointed out in Ref. \cite{YangYao}, where a correlation length was stated. In the following, we will first  present Monte-Carlo data on correlation functions for completeness, and 
then turn to the study of entanglement properties of the square-lattice RVB state \Eq{state}.

\subsection{Correlations}
We begin by investigating the connected dimer-dimer (i.e., singlet-singlet) correlation functions, both for the original RVB wave-function \eqref{state} as well as for its projections onto individual topological sectors.  Such projections can be achieved by means of operations that flip the pre-Kasteleyn orientation $\tilde\Theta$ along non-trivial loops, which change the relative signs between topological sectors.
Note that in the preceding section, we have defined topological sectors through the dynamics associated with (arbitrarily large) contractible resonance loop moves. Alternatively, one may define them in terms of two ``winding numbers'' \cite{ReadChakraborty89} that each count the parity of the number of dimers crossed by one of two fixed non-contractible loops (which together generate the fundamental group of the torus). This defines precisely four topological sectors, all of which are invariant under the resonance-move dynamics of the first definition. To the extent that these dynamics are ergodic in each sector, the two definitions are identical. It is not difficult to argue that this is the case, since the loops for the resonance moves considered need only be contractible and are otherwise arbitrary. We will now denote the winding numbers associated along two non-contractible loop, on along the $x$-direction and on along the $y$-direction, and going through the crosses of the frustrated square lattice, by $n_{x}$ and $n_{y}$, respectively. We denote their possible values by $e$ (even) and $o$ (odd), and thus have the notation $(n_{x},n_{y})=(e,e), (o,e), (e,o), (o,o)$ for the sectors, which we will also abbreviate through an index $i$ running from $1$ through $4$, writing $\ket{\Psi_i}$ for the projections of the RVB state onto the topological sectors. The winding numbers $n_{x}$, $n_{y}$ correspond to 't Hooft magnetic loop operators in Ising gauge theories, which are readily associated to quantum dimer models through well-studied mappings \cite{MoessnerSondhiFradkin, Fradkin_book}, though the connection with RVB wave functions is, of course, somewhat less immediate. 

\begin{figure}[tbp]
\centering
\mbox{\subfigure{\includegraphics[width=3.5in]{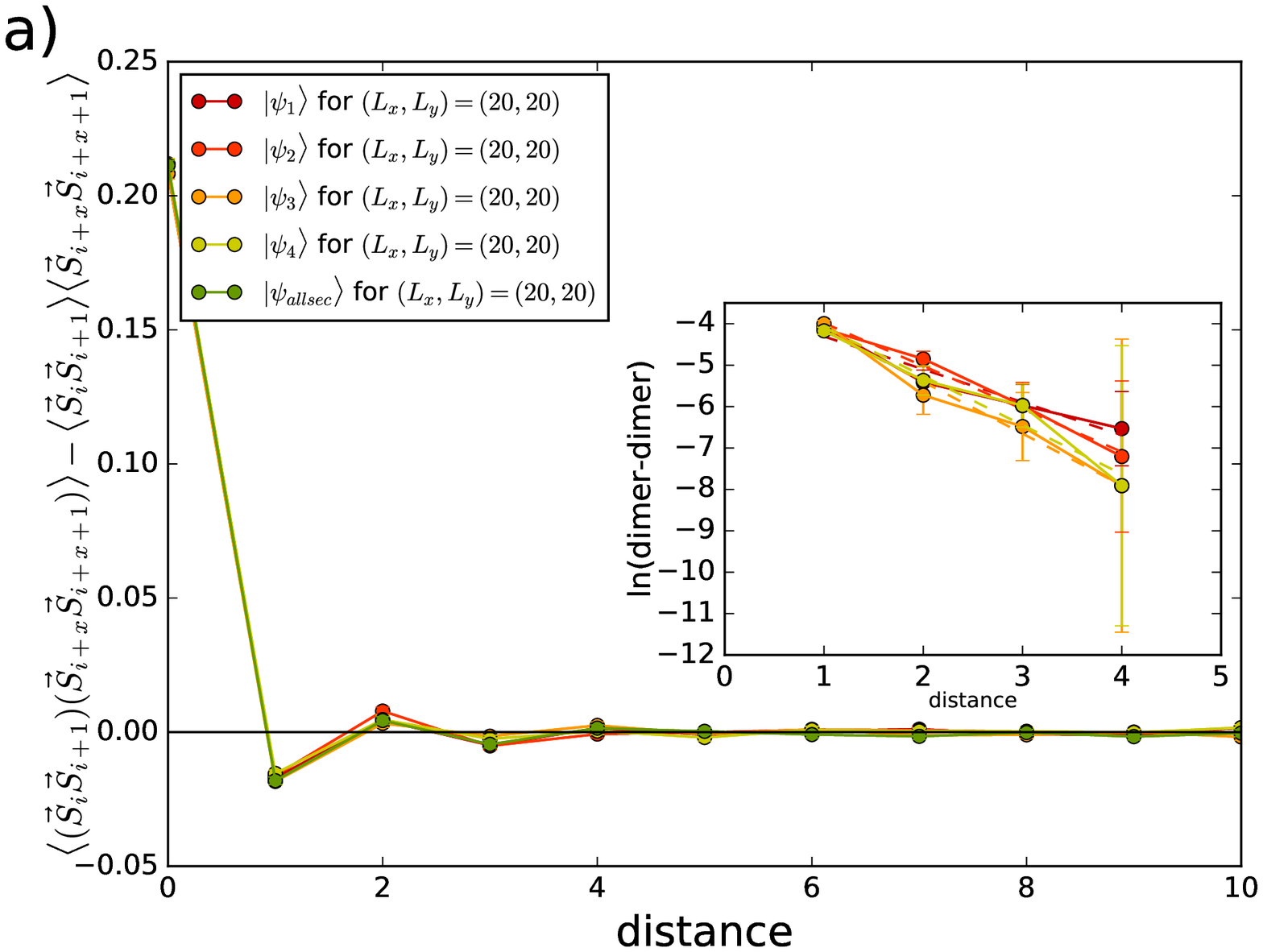}}\quad \quad 
\subfigure{\includegraphics[width=3.5in]{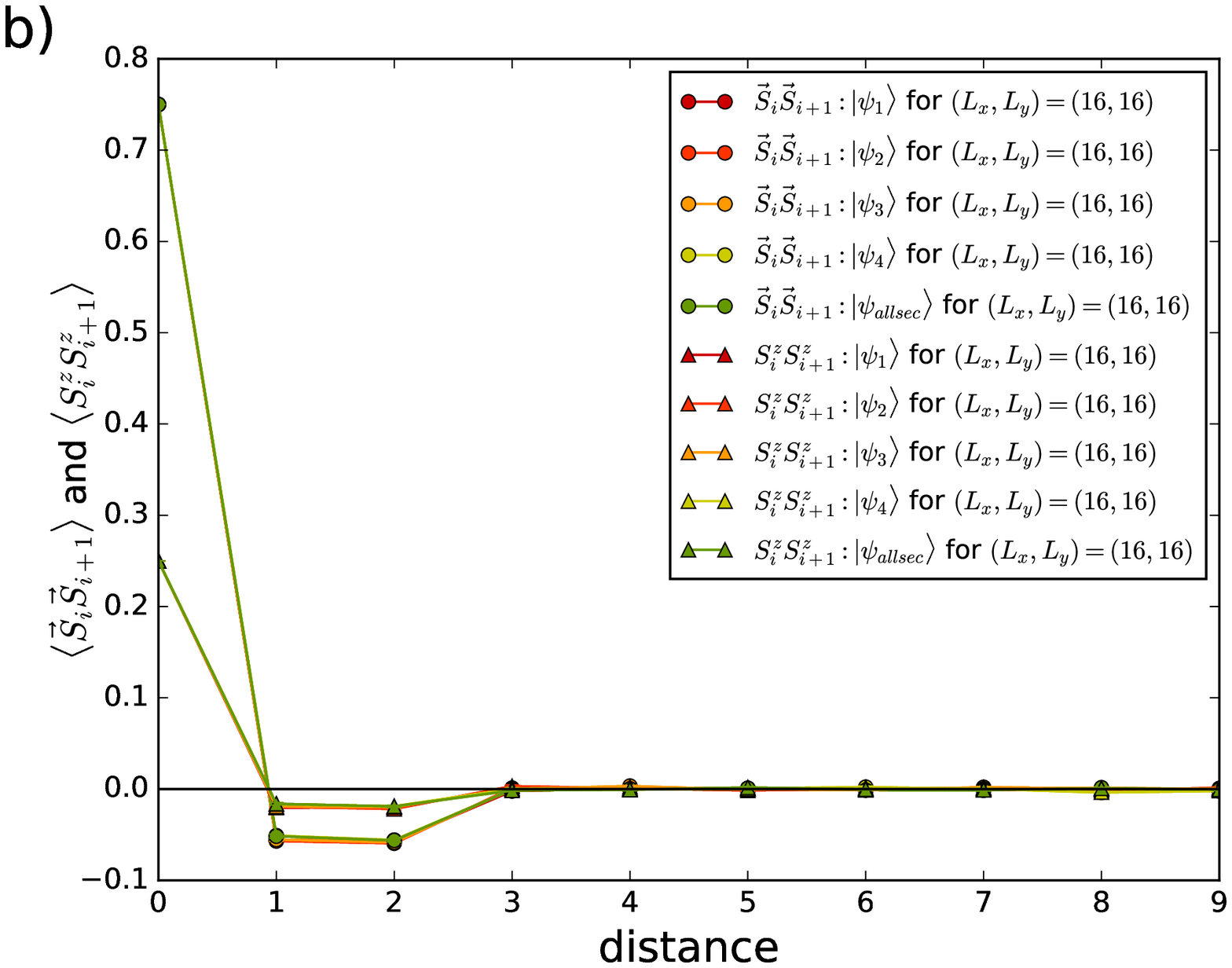} }} 
\caption{Correlation functions of various local operators. 
a) The connected singlet-singletcorrelation 
$\langle (\vec{S}_{i} \vec{S}_{i+1})(\vec{S}_{i+\kappa} \vec{S}_{i+1+\kappa})\rangle - \langle \vec{S}_{i} \vec{S}_{i+1}\rangle \langle \vec{S}_{i+\kappa} \vec{S}_{i+1+\kappa}\rangle$ for a system of size $(L_{x}, L_{y}) = (20,20)$ within a topological 
sector $|\Psi_{i}\rangle$, $i = 1,\ldots,4$. We observe that the correlations are the same within error bars. Thus, as for the local expectation values of Fig. \ref{average_heisenberg}, the four topological sectors to these local correlators. The insets demonstrates exponential decay on a logarithmic scale.
b) The correlation functions $\langle \vec{S}_{i} \vec{S}_{i+\kappa}\rangle$ and 
$\langle \vec{S}^{z}_{i} \vec{S}^{z}_{i+\kappa}\rangle$ for each topological sector. Again, the correlations decay exponentially and they are 
indistingushible within errors. SU(2)-invariance, while not manifest in the Pfaffian-formulation, is restored in the correlators, as  
we numerically confirm $\langle \vec{S}_{i} \vec{S}_{i+\kappa} \rangle = 3 \times \langle \vec{S}_{i}^{\alpha} \vec{S}_{i+\kappa}^{\alpha} \rangle$ with $\alpha = x,y,z$. 
} \label{figure_3}
\end{figure}

 Figure \ref{figure_3} shows that the dimer-dimer correlations 
are indistinguishable within error bars, both for the equal amplitude superposition over all topological sectors \Eq{state}, as well as
for the projections onto any one of the four topological sectors. 
A linear fit shows that indeed correlations fall off exponentially, as expected for a gapped $\mathbb{Z}_{2}$ spin liquid. 
We also computed the correlations $\langle \vec{S}_{i}\cdot \vec{S}_{i+\kappa} \rangle$ and 
$\langle \vec{S}_{i}^{z} \vec{S}_{i+\kappa}^{z} \rangle$. Again, all correlations decay exponentially, topological sectors 
remain indistinguishable. Additionally, we checked that 
$\langle \vec{S}_{i} \vec{S}_{i+\kappa} \rangle = 3 \times \langle \vec{S}_{i}^{\alpha} \vec{S}_{i+\kappa}^{\alpha} \rangle$ with 
$\alpha = x,y,z$. This is an obvious consequence of $SU(2)$-invariance, which, however, is not manifest in the Pfaffian Ising-basis formulation, and thus serves  as a consistency check. 
To complete the investigation of the behavior of local operators,  we have measured the expectation value 
$\langle \vec{S}_{i} \vec{S}_{i+1} \rangle$ along the horizontal, vertical, and diagonal bonds. 
For each type, measured values were indistinguishable within error bars between different topological sectors for the larges systems sizes studied (Fig. \ref{average_heisenberg} and Table \ref{tab:table0}). Indistinguishable topological sectors are of a piece with the unbroken symmetry of the Hamiltonian in the ground states, and the topological nature of the ground state degeneracy. 

\subsection{Entanglement}

After verifying the exponential decay of the correlations, we now turn our attention to computing  topological properties. It is now well established that much interesting information about the universal characteristics of any physical phase are encoded in entanglement properties. This is true for topological phases \cite{lw,KP}, holographic quantum matter \cite{hartnoll}, and many others.
In the case of bipartite entanglement, where the lattice is divided into a region $A$ and its complement $B$, one may consider the 
 Renyi entropy of order $n$ defined as 
$S_{n} = \text{ln Tr}(\rho_{A}^{n})/(1-n)$, 
where $\rho_{A} = \text{Tr}_{B}|\Psi \rangle \langle \Psi|$ is the reduced density matrix of region $A$. 
Ground states of  gapped local Hamiltonians are known to exhibit an {\it area law} scaling in region size, 
which in two dimensions can generically be written as, 
$S_n(\rho_{A}) = \alpha_n L_{A} - \gamma + \cdots $ \cite{Eisert}. 
Here, the leading term is dependent on the ``area'' (or boundary) of region $A$. 
The second term, the topological entanglement entropy (TEE) $-\gamma$ \cite{lw,KP}, 
is characterized by the total quantum dimension $\mathcal{D}$, which is defined through the
quantum dimensions of the individual quasiparticles $d_{i}$ of the underlying theory: 
$\mathcal{D} = \sqrt{\sum_{i} d_{i}^2}$ \cite{lw,KP,Dong}.
Conventionally ordered phases have $\mathcal{D} = 1$,
while topologically ordered phases have $\mathcal{D} > 1$ with the
TEE given by $\gamma = \text{ln}(\mathcal{D})$.
The TEE is a fingerprint of the underlying topological phase, but it does not in general uniquely identify the topological phase. 

 For sufficiently large areas, $\gamma$ is a universal value independent of the shape of the regions involved
 {\em as long as} the area $A$ is {\em contractible}. 
However, in the case where the area $A$ has at least one non-contractible boundary 
the TEE $\gamma$ becomes state-dependent, i.e., is sensitive the  specific linear combination of degenerate ground states it is evaluated for.
As shown in Ref.~\cite{zhang}, further developing ideas of Ref. ~\cite{Dong}, 
if one expresses 
any ground state in the basis of the {\it minimum entropy states} (MES-states), 
$| \Psi_{\alpha} \rangle = \sum_{j} c_{j} | \Xi_{j} \rangle$, 
then the sub-leading constant to the area law from cut of the torus into two cylinders is 
\begin{eqnarray}\label{gamma_torus2} 
\gamma^{\prime}(\{p_{j}\}) = 2\gamma + \text{ln}\bigg(\sum_{j} \frac{p_{j}^{2}}{d_{j}^{2}} \bigg) \;. 
\end{eqnarray}
for $S_{2}$, where $p_{j} = |c_{j}|^2$. We further discuss MES-states in the results to follow. 

\begin{figure}[tbp]
\centering
\mbox{\subfigure{\includegraphics[width=3.5in]{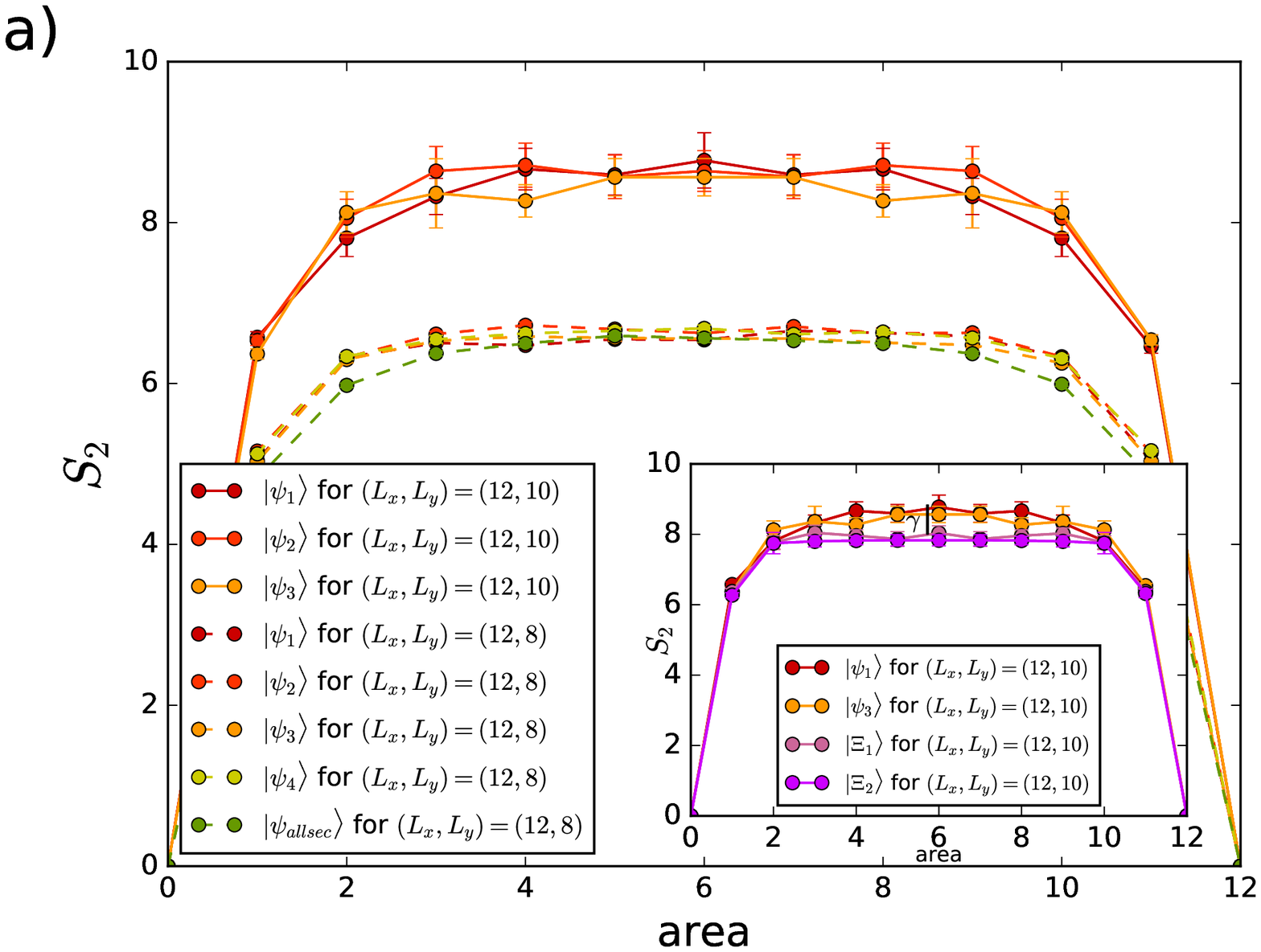}}\quad \quad
\subfigure{\includegraphics[width=3.5in]{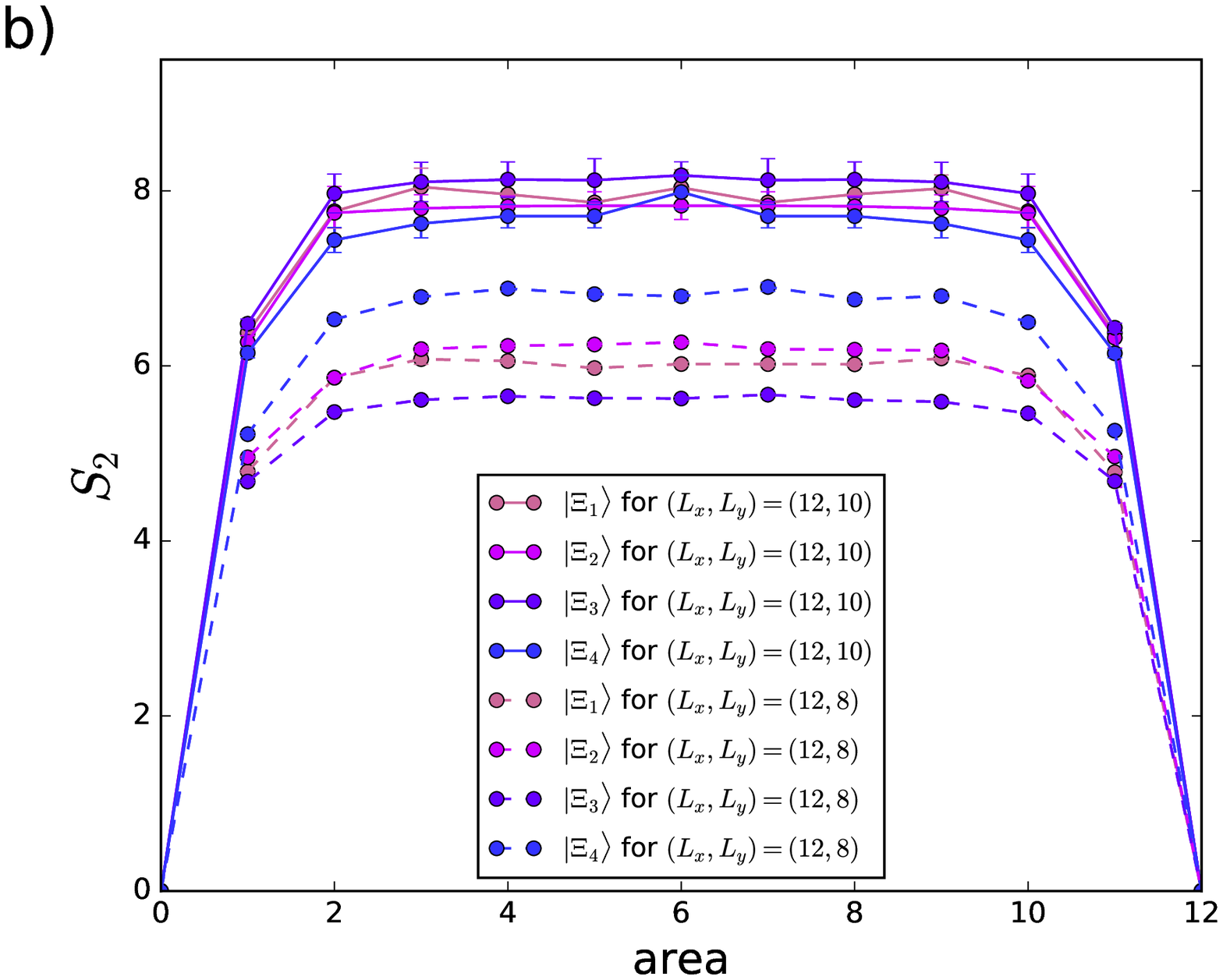} }}
\caption{Renyi entropies $S_2$ for cylindrical cuts of the torus and various system sizes, topological sectors, and linear combinations thereof. a) $S_{2}$ as a function of cylindrical height for all topological sectors
and their equal amplitude superposition.
All data agree well for system size 96, in agreement with the MES Ansatz \Eq{MES_Z2}, though small discrepancies can still be resolved. These are interpreted as finite size effects.
Error bars are necessarily larger for larger system size, as the expectation value of the SWAP operator \cite{hastings,RatioT} becomes exponentially small.
b) $S_2$ for the four minimum entropy states (MES) $|\Xi_{i}\rangle$ according to the Ansatz \Eq{MES_Z2}. Stronger discrepancies between the different MES-states are noted at smaller system size, but agreement within error bars is again attained for 120 sites.
The inset in a) illustrates the procedure from which value for the topological entanglement entropy
is extracted via the difference in $S_2$ for MES-states and the $\ket{\psi_i}$. See text for details.
} \label{figure_5}
\end{figure}
In the following, we investigate entanglement properties of the frustrated spin-$\frac{1}{2}$ RVB 
wave functions using the variational Pfaffian MC scheme for lattices of up to $120$ sites.
As mentioned above, the Pfaffian MC scheme allows one to project onto each topological sector, and every linear combination thereof.
We will again use this feature in the following results.
To obtain the second Renyi entropy $S_{2}$ for noncontractible regions, 
we employ the standard QMC replica-trick \cite{hastings,RatioT}. 

We have  calculated the TEE for cylindrical regions $A$ that are wrapped around the torus like ``ribbons'', and whose boundaries thus correspond to generators of the fundamental group of the torus. Within our method, this approach has proven advantageous compared to schemes extracting the TEE from contractible regions, 
such as the Levin-Wen \cite{lw} and the Kitaev-Preskill \cite{KP} constructions. These constructions isolate the TEE from the entanglement $S_{2}$ 
by canceling the area dependent contributions when $S_{2}$ is computed for several suitably chosen areas. However, these procedures usually 
lead to values for the TEE $\gamma$ that are plaqued by rather large error bars. 
For better accuracy, we only consider regions with noncontractible boundaries. 
We examine frustrated RVB states for square lattices  of dimensions $12\times 8$ and $12 \times 10$, respectively and calculate the Renyi entropy $S_{2}$ for cylindrical bipartitions. 
As the cylinder length increases, $S_{2}$ quickly saturates (Figure \ref{figure_5}a).
This type of behavior is consistent with the system having a gap. 

For the smaller lattice size of 96, Figure \ref{figure_5}a) shows the $S_{2}$-values of different topological sectors to be in near agreement, despite some small discrepancies that are nonetheless resolved by the very small error bars . We interpret these discrepancies as finite size effects. While \Eq{gamma_torus2} predicts discrepancies in TEE for {\em some} linear combinations of the MES-states, the expected topological phase in question has $d_{i}=1$ for all $i$. Therefore, for the MES-states themselves, all TEE values are expected to be degenerate. The same will be true for any other ground state basis with coefficients in the MES-basis whose absolute values differ from one basis state to another only by permutation. 
The data for 120 lattice sites do not resolve any discrepancy in $S_{2}$ for different topological sectors within the (somewhat larger) error bars. We will argue any such discrepancies  
to be due to finite size effects and error bars, as we will explain in the following.


Our discussion will be based on an  informed guess for the MES-states. 
As mentioned above,  at least for hard-core dimer versions 
of our wave functions to be discussed in the following Section, the topological sectors $(n_{x},n_{y})$ 
can be identified with 't Hooft magnetic loop eigenspaces of  Ising gauge theory, or, for that matter, the toric code. For the latter, the linear relations between the ground states carrying well-defined 
't Hooft magnetic loop quantum numbers and the MES-ground-states is exactly known \cite{zhang}. At least for the kagome quantum dimer model of \cite{misguich}, the same calculation goes through \cite{wildeboer15}, yielding the same relations. One expects these relations to be universal for the  $\mathbb{Z}_{2}$ topological phase. Thus, as long as the 
't Hooft magnetic loop quantum number carrying ground states can be unambiguously identified, the MES-states are known. For the RVB-states, this is the case, as long as we can lift these quantum numbers from the hard-core dimer picture associated to these wave functions. 

Based on the above assumptions, one arrives at the following Ansatz for MES-states associated with cuts along the $y$-direction: 
\begin{equation}
| \Xi_{1,2}\rangle = \frac{1}{\sqrt{2}} (| \Psi_{ee} \rangle \pm | \Psi_{eo} \rangle) , \qquad 
| \Xi_{3,4}\rangle = \frac{1}{\sqrt{2}} (| \Psi_{oe} \rangle \pm | \Psi_{oo} \rangle)\;. 
\label{MES_Z2}
\end{equation}
Note that phase ambiguities regarding the $\ket{\Psi_{n_x,n_y}}$ are essentially eliminated by time-reversal symmetry, i.e., the requirement of working with ``real'' wave functions. Note also that the inverse relations 
\begin{equation}\label{singlesec}
| \psi_{ee(eo)}\rangle = \frac{1}{\sqrt{2}} (| \Xi_{1} \rangle \pm | \Xi_{2} \rangle) , \qquad
| \psi_{oe(oo)}\rangle = \frac{1}{\sqrt{2}} (| \Xi_{3} \rangle \pm | \Xi_{4} \rangle)\;. 
\end{equation}
indeed give the same TEE $\gamma'_{\sf magnetic}=\gamma$ for $d_{i}\equiv 1$ in \Eq{gamma_torus2}. 
While the above relations are based on some assumptions, in particular, that the RVB-states do indeed belong to the $\mathbb{Z}_{2}$ topological phase, a strong consistency requirement is given by the fact that when $\gamma$ is now calculated based on the above identifications of MES-states, the value
$\ln(2)\approx 0.69$ must be obtained. 

To this end, we compare Monte-Carlo results for $\gamma'$ for both the ``magnetic'' eigenstates 
$\ket{\Psi_{n_x,n_y}}$ (Figure \ref{figure_5}a) and the conjectured MES-states $\ket{\Xi_i}$ (Figure \ref{figure_5}b). 
We first observe a pronounced reduction in the $S_2$ discrepancies between different $\ket{\Xi_i}$  
when comparing data for 96 and 120 lattice sites. This is again consistent with these discrepancies being due to finite system size. 
Indeed, from \Eq{gamma_torus2}, we expect $\gamma'_{\sf MES}=2\gamma$ for all MES-states. We now proceed by evaluating the {\em difference} in $S_2$ between the $\ket{\Xi_{i}}$ and the $\ket{\Psi_{n_x,n_y}}$ for the size 120 lattice,
which consistency requires to be equal to $\gamma'_{\sf MES}-\gamma_{\sf magnetic}=2\gamma-\gamma=\gamma$. 
In the following, we extract this value by averaging 
 this difference over cylinder lengths $x = 5, \ldots, 7$, and moreover, we may average over different topological sectors and MES-states, respectively. For example, working with $\ket{\Xi_{1}}$ and $\ket{\Xi_{2}}$ their linear combinations $\ket{\Psi_{ee}}$ and $\ket{\Psi_{eo}}$, respectively, one extracts six independent measurements of $\gamma$ with average $0.73$ and standard deviation $0.06$. 
The measurement of the logarithm of an exponentially small value is highly demanding. We can improve the accuracy by replacing $S_2$-values from the $\ket{\Psi_{eo}}$-sector with the values for $\ket{\Psi_{oe}}$, which are, at the time of this writing, better converged. The resulting average over six values and their standard deviation are then found to be $0.70\pm0.04$. 
There is thus encouraging  agreement with the expected value of $\ln(2)$, indicating the consistency of our assumptions in identifying the MES-states. Another consistency check lies in the fact that the equal amplitude superposition over all $\ket{\Psi_i}$ has essentially the same $S_2$ value as the individual $\ket{\Psi_i}$ for the $96$ site lattice (Figure \ref{figure_5}a), as implied by Eqs. \eqref{singlesec} and \eqref{gamma_torus2}. Overall, these data well support the hypothesis that the wave function \eqref{state} describes a state in the $\mathbb{Z}_2$ topological phase.

\section{A Quantum-Dimer Model\label{QDMSEC}}
\subsection{General Hamiltonian} 

In Sec. \ref{RVBdef} we emphasized that the definition of the RVB-state \eqref{state} 
utilizes only local rules. Though we will not systematically elaborate here, it is likely that 
the \Eq{state} can be written as a tensor-network state, as have other short-range RVB wave functions \cite{schuch}. Related to that, it can be given a parent Hamiltonian comprised of local operators acting on sufficiently large cells, following the recipe of Ref.\cite{seidel09} or equivalently \cite{zhou} Ref.\cite{schuch}. Here we will chose the route of quantum dimer models \cite{Kivelson89} to motivate that the amplitudes in \Eq{state} may arise naturally from a local Hamiltonian at least in this context.

To this end, we will now simplify the Hilbert space of the original spin-$1/2$ degrees of freedom of \eqref{state} according to the traditional quantum dimer model philosophy \cite{Kivelson89}. 
The quantum dimer model is defined on a Hilbert space with distinct, orthonormal states corresponding to each allowed  dimer 
covering of the lattice, in our case, the frustrated square lattice. While representing singlets in an abstract sense, these dimers are more properly understood as hard-core bosons living on the medial lattice. In particular, different hard-core boson states have different inner products from the valence bond states associated with the same dimerizations. In the following, we will re-interpret the kets $\ket{D}$ 
as the hard-core dimer configuration associated with the dimerization $D$, and \underline{not} the valence bond configuration associated to $\ket{D}$ in preceding sections. 
We define a quantum dimer model (QDM) Hamiltonian via matrix elements between at most locally differing dimer states. 
There are two different types of matrix elements: the ``potential'' terms, 
in the following labeled $V,\ V^{\prime},\ V^{\prime\prime}$ and $V^{\prime\prime\prime}$, which are diagonal in the dimer basis and 
associate an interaction energy with various local arrangements of dimers, and  ``kinetic'' terms, in the following referred to as 
$t,\ t^{\prime}, t^{\prime\prime}$, and $t^{\prime\prime\prime}$, 
which involve a local rearrangement of a small number of dimers. 
On the nonbipartite square lattice, we represent the Hamiltonian graphically as: 
\begin{eqnarray}\label{eq:square}  
{\cal H}&=&\sum_{\square}\Bigg[\!-\!t\left(\ket{\input{12-34.latex}} \bra{\input{13-24.latex}} 
\!-\!\frac{1}{2}\ket{\input{12-34.latex}} \bra{\input{14-23_julia.latex}} 
\!-\!\frac{1}{2}\ket{\input{13-24.latex}} \bra{\input{14-23_julia.latex}} \!+\!h.c.
\right) 
\!+\!V\left(\frac{3}{2}\ket{\input{12-34.latex}}\bra{\input{12-34.latex}} + \frac{3}{2}\ket{\input{13-24.latex}}\bra{\input{13-24.latex}} 
+ \ket{\input{14-23_julia.latex}}\bra{\input{14-23_julia.latex}}   \right)\Bigg]\nonumber
\\
\!&+&\!\sum_{\{\input{rect.latex}\}} 
\Bigg[\!-\!t^{\prime}\left(\ket{\input{13-45.latex}}\bra{\input{14-35.latex}} +h.c.\right)\!+\!V^{\prime}\left(\ket{\input{13-45.latex}} \bra{\input{13-45.latex}} +\ket{\input{14-35.latex}}\bra{\input{14-35.latex}}\right)\!-\!
t^{\prime\prime}\left(\ket{\input{13-46.latex}} \bra{\input{14-36.latex}} \!+\!h.c.\right)\!+\! V^{\prime\prime}\left(\ket{\input{13-46.latex}}\bra{\input{13-46.latex}} \!+\!
\ket{\input{14-36.latex}}\bra{\input{14-36.latex}}\right)
\Bigg] \nonumber
\\
\!&+&\!\sum_{\{\input{rectt.latex}\}} 
\Bigg[\!+\!t^{\prime\prime\prime}\left(\ket{\input{14-36_julia.latex}}\bra{\input{14-36_julia3.latex}} +h.c.\right)\!+\!V^{\prime\prime\prime}\left(\ket{\input{14-36_julia.latex}} \bra{\input{14-36_julia.latex}} +\ket{\input{14-36_julia3.latex}}\bra{\input{14-36_julia3.latex}}\right)  
\Bigg], 
\end{eqnarray}
where black bonds represent dimers. In \eqref{eq:square}, the first sum runs over all plaquettes, 
and the second sum runs over all pairs of adjacent plaquettes (with both orientations) while the third sum runs
over all ``3-by-3''-plaquettes. For the ``3-by-3''-plaquettes the braces symbolize that we take into account 
all four possible orientations that the dimer emanating from the center site can have while occupying a diagonal link. 
The circumstance that, somewhat atypically, kinetic terms appear with positive sign in some places accounts
for the fact that the ground state to be stabilized, i.e., \Eq{state} re-interpreted as quantum dimer wave function, has negative amplitudes associated with crosses. 
The model \eqref{eq:square} contains kinetic and potential terms that involve dimers living on diagonal links and 
pairs of crossed dimers. 
It is thus similar to the one previously investigated by 
Yao and Kivelson \cite{yao_kivelson}, but expands the latter by including configurations and matrix elements involving crossed pairs of dimers. 

We emphasize that despite similarities in notation and terminology, QDM differ decisively from spin-1/2 systems. Whether the  exploration of the phase diagram of QDMs yields actual insights into that of $SU(2)$-invariant spin-1/2 systems on the same lattice has been of great interest since the original QDM of Rokhsar and Kivelson \cite{Kivelson89}. In particular,  RVB wave functions of the form \Eq{RVBp} have been studied extensively both for bipartite \cite{alet, tang} and non-bipartite lattices \cite{wildeboer12, schuch, YangYao, wildeboer15}, and have always been found to have qualitatively similar features as their QDM counterparts.   The present case of interest, \Eq{state}, may be viewed as one step further removed from the original RVB-state when regarded as a quantum dimer wave function. Indeed, while in many lattice geometries with planar graphs, the valence bond state for this graph are now well-known to be linearly independent \cite{wildeboer11}, this is not so for non-planar case discussed here. In particular, crossed pairs of valence bonds and the two parallel configurations on the same square are linearly dependent: 
$|
\input{14-23.latex}
\rangle
\propto -(
|
\input{12-34.latex}
\rangle 
+
|
\input{13-24.latex}
\rangle
)$. 
(It is currently an unknown problem whether the valence bond configurations discussed here exhibit further linear dependences other than the aforementioned for large lattices.) Linear independence has often been employed to strengthen the case of correspondence between QDMs and frustrated $SU(2)$-invariant quantum antiferromagnets, e.g., by means of systematic derivation of QDMs from antiferromagets via formal expansion in the ``overlap parameter'' \cite{schwandt}, which characterizes non-orthogonality between different valence bond configurations. Such expansions are well-defined only when linear independence is given. 
In the context of the present study, it is natural to extend questions of general correspondence between QDMs and anti-ferromagnets and/or their prototypical RVB-type trial states to situations where the linear independence is absent.

\subsection{The RK point.} \label{rk_point_sec} 
The parameter space of the model \eqref{eq:square} has a ``generalized Rokhsar-Kivelson'' line where the exact ground states is know. In the following, we specialize to this line defined by the following choce of parameters: 
\begin{eqnarray}
t=V, ~t'=V', t^{\prime\prime}=V^{\prime\prime} ~\textrm{and}~ 
t^{\prime\prime\prime}=V^{\prime\prime\prime} \;\;\mbox{(all couplings positive)}\;.
\end{eqnarray}
One can show that at this generalized RK line, \eqref{eq:square} can be expressed as a sum over positive semi-definite 
operators  with ground state(s) formally analogous to \eqref{state}, which we repeat here as 
\begin{eqnarray}\label{eq:rk}
\ket{\Psi}=\sum_{D}(-1)^{n_{c}} \ket{D} \;\;\mbox{(quantum dimers)},
\end{eqnarray}
emphasizing that this is {\em not} the same as the RVB wave function \eqref{state}, but rather its quantum dimer version. As in the RVB case, for periodic boundary conditions four topological sectors may be introduced. As the dynamics of the model do not mix these sectors, the sum over $D$ may be restricted to any one sector, giving four ground states on the torus.
It is indeed not difficult to see that each of the four types of local terms  in \Eq{eq:square}, 
proportional to one of the fours independent couplings at the RK line, individually annihilates 
\Eq{eq:rk}. The quantum dimer RK state, and its projections onto topological sectors, thus simultaneously minimize each of these local terms.

One expects \Eq{eq:rk} to describe a $\mathbb{Z}_2$ topological liquid just as the original 
RVB state \Eq{state} does, based on the results of the preceding section. Consider the dimer-dimer 
correlation fuction $\langle \Psi|n_i n_j|\Psi\rangle/\langle\Psi|\Psi\rangle$, where $n_i$ is the dimer occupation operator on a link labeled $i$. It is easy to see that this does map onto a classical dimer problem, and so can be addressed by Monte-Carlo evaluation. We will leave this to future studies. 
The point we wish to make here, however, is that this classical problem is {\em oblivious} to the phases in \Eq{eq:rk}. It is therefore not addressable through the Pfaffian methods that have been so successful in solving many classical dimer problems, and their quantum dimer analogues. 
Specifically, the Pfaffian formulation based on the pre-Kasteleyn orientation of the preceding section {\em does} always reproduce the phases of \Eq{eq:rk}, as we have seen, and therefore cannot produce the infinite temperature classical partition function that effectively describes the dimer-dimer correlator.
We leave the evaluation of correlation functions of the QDM-state \eqref{eq:rk} as an interesting problem for the future. We point out that that the pre-Kasteleyn orientation discussed here has been previously utilized by Yao and Kivelson to gain analytic insight into correlations of a crossing-free variant of the QDM on the frustrated square lattice\cite{yao_kivelson}. Again, exact Pfaffian evaluation is not possible, but an exact reformulation as an {\em interacting} fermionic theory allows controlled mapping onto an effective massive Thierring model. We do not rule out that variants of local lattice Hamiltonians exists whose correlators reduce exactly to Pfaffians of pre-Kasteleyn matrices. We leave this as another interesting problem for the future.

\section{Conclusion\label{conclusion}} 

In this work, we have formally defined the notion of a pre-Kasteleyn orientation. A special instance on the frustrated square lattice, which had already appeared in the study a ``crossing-free'' quantum dimer models \cite{yao_kivelson}, has been utilized for a Pfaffian re-formulation of a next-nearest neighbor spin-1/2 RVB wave function {\em with} crossings and a negative sign rule on the square lattice. This allowed us to evaluate the $S_{2}$-Renyi entropy on the torus using Pfaffian Monte-Carlo \cite{wildeboer12, wildeboer15}, confirming the wave functions ``minimum entropy state'' \cite{zhang} by consistent extraction of topological entanglement entropy of $\ln(2)$. This corroborates the fact that this wave function lies in the $\mathbb{Z}_2$ topological phase \cite{YangYao}.  In addition, local expectation values, their dependence on topological sector, and correlation functions have been evaluated in some detail. As a by-product, various useful statements surrounding pre-Kasteleyn orientations were proven, which in particular address technicalities from the possibility of (self-)intersecting loops in the overlap graph between two different dimerizations. We also discussed a quantum dimer model related to the RVB wave function. We conjecture that pre-Kasteleyn orientations may prove to have broader applications facilitating combined analytic-numerical techniques for lattice wave functions defined on non-planar graphs, and are hopeful that our work will stimulate future investigation.

\begin{acknowledgments}
JW thanks NSF DMR 1306897 and NSF DMR 1056536  for partial support. AS gratefully acknowledges insightful discussions with K. Yang and Z. Nussinov. 
Our Monte Carlo codes are partially based upon the ALPS libraries \cite{alps1, alps2}. The simulations were run on the SHARCNET clusters. 
\end{acknowledgments}

\appendix 
\section{Kasteleyn-like properties of various lattice graphs} \label{app}

In this short appendix, we first show the equivalence of pre-Kasteleyn and Kasteleyn for lattices with planar topology stated in Sec. \ref{MCmethod}.
 Subsequently, we will explain why the non-planar frustrated square lattice equipped with the orientation shown in Fig. \ref{figure_1}b) is pre-Kasteleyn.

\begin{figure}[tbp]
\includegraphics[width=0.4\textwidth]{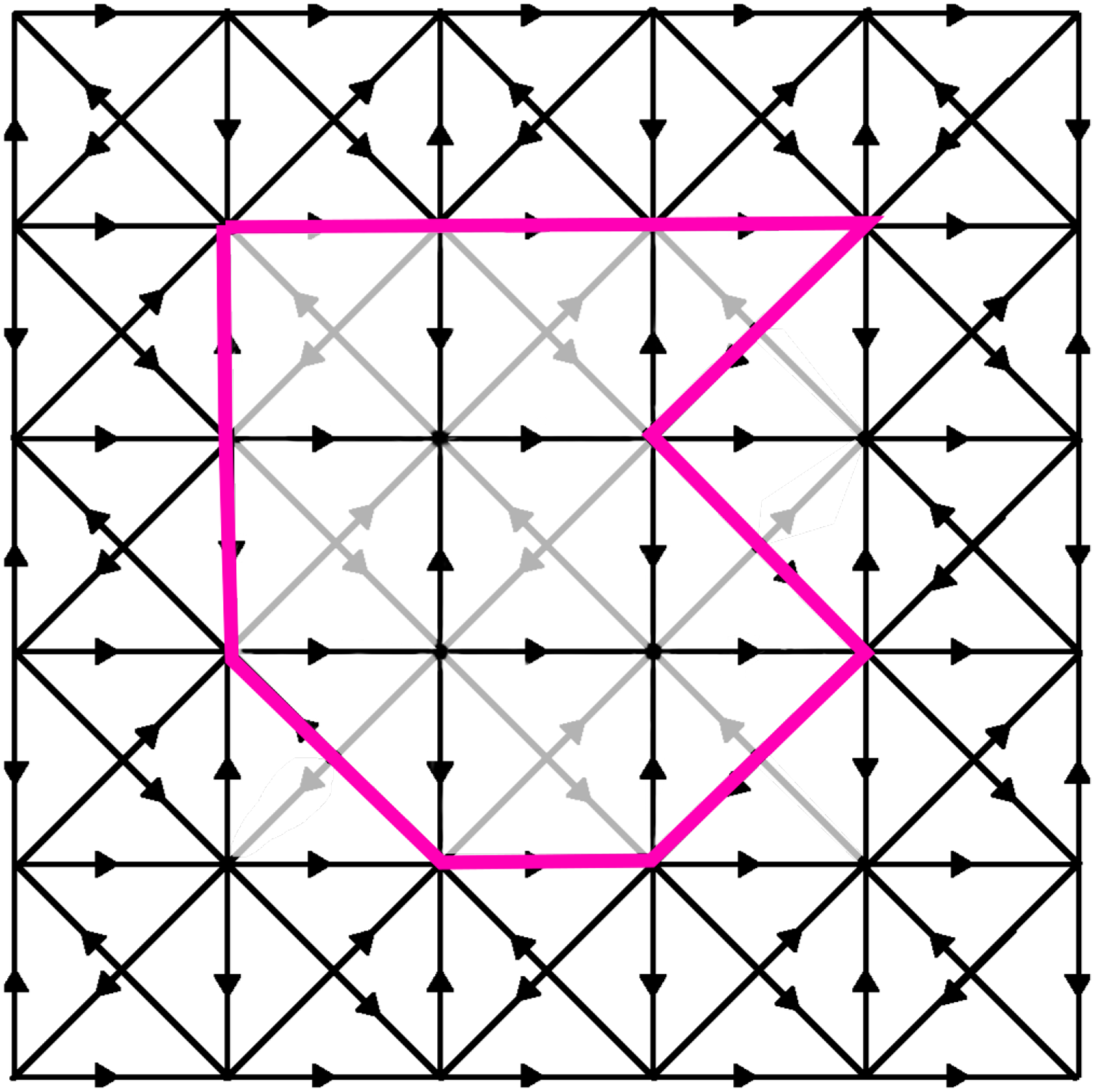}
\caption{
A closed non-intersecting loop on the frustrated square lattice. The orientation shown is the pre-Kasteleyn orientation of Fig. \ref{figure_1}b). After removal of certain diagonal links as specified in the text (light gray), the loop borders a planar Kasteleyn-oriented graph. 
}
\label{loop}
\end{figure}

For planar lattice topology, clearly, pre-Kasteleyn implies Kasteleyn, simply by applying the defining property of the former to loops around faces. 
For the opposite implication, consider any closed loop for a planar lattice topology equipped with a Kasteleyn orientation $\tilde\Theta$. The loop defines an oriented planar sub-graph $G$ that has the loop as its boundary. (A boundary edge belongs to one face only, we never include the ``outer face'' of a planar graph among its faces.) Being planar, the Euler index of $G$ is $1$, thus
\be
       V-E+F=1,
\ee
where $V$ is the number of vertices, $E$ the number of edges, and $F$ the number of faces of $G$.
(For the following considerations about $G$, although Fig. \ref{loop} shows a loop on the frustrated square lattice, it may also serve to illustrate the present situation, if light gray links are ignored.)
If $f$ denotes a face of $G$, let $E_c(f)$ be the number of clockwise oriented edges of $f$. 
In $\sum_f E_c(f)$, each internal edge of G, i.e., each edge that is {\em not} at the boundary, is counted exactly once, since for the two faces it borders, it will count as clockwise for precisely once.
Therefore,
\be
\sum_f E_c(f)= E_{c,b}+ E_{\sf in}\,,
\ee
where $E_{c,b}$ is the number of clockwise edges on the boundary, and $E_{\sf in}$ is the number of internal edges. On the other hand, for a Kasteleyn orientation, each $E_c(f)$ has to be odd, thus
\be
\sum_f E_c(f)= F  \mod 2\,.
\ee

Together, the last three equations give
\be\label{a4}
   V+E_{c,b}+ E_{\sf in}-E=1 \mod 2\,.
\ee
Noting that $E-E_{\sf in}=E_b$, the number of boundary edge, which is the length of the boundary loop, and thus also equal to the number of boundary vertices $V_b$, \Eq{a4} gives $V_{\sf in}+E_{c,b}=1\mod 2$, where $V_{\sf in}=V-V_b$ is the number of vertices enclosed by the loop.
As the sign of an integer is irrelevant modulo 2, this can finally be expressed as
\be\label{preK}
     V_{\sf in}=E_{c,b}+1 \mod 2 \, ,  
\ee
which proves the assertion.

Last, consider now a closed loop on the frustrated square lattice topology equipped with the orientation $\tilde\Theta$, as in Fig. \ref{loop}.
It is easy to see that if we remove all diagonal edges from the lattice that have at at least one vertex in the interior of the loop ({\em not} including the loop itself!), the resulting subgraph $G$ consisting of the loop and its interior is now a planar Kasteleyn graph. Therefore, \Eq{preK} again applies. 

More generally, any lattice with an oriented link topology is pre-Kasteleyn if we can establish that for any closed, non-intersecting loop, we can obtain a Kasteleyn graph (in the ordinary sense) by means of removal of links that do not belong to the loop. While, based on the results above, this criterion is quite trivially equivalent to the pre-Kasteleyn property, it is quite useful if not too many links, or only certain types of links, need to be removed (at least inside the loop), as in the present case.

\end{document}

%% file: 12-34.latex
%
%
\setlength{\unitlength}{3947sp}
\begin{picture}(160,120)(-20,10)
 \put (0,0){\circle*{25}}
 \put (120,0){\circle*{25}}
 \put (0,120){\circle*{25}}
 \put (120,120){\circle*{25}}
 
 \put (0,10){\circle*{1}}
 \put (0,20){\circle*{1}}
 \put (0,30){\circle*{1}}
 \put (0,40){\circle*{1}}    
 \put (0,50){\circle*{1}}
 \put (0,60){\circle*{1}}  
 \put (0,70){\circle*{1}}
 \put (0,80){\circle*{1}}    
 \put (0,90){\circle*{1}}
 \put (0,100){\circle*{1}}    
 \put (0,110){\circle*{1}}
 
 \put (120,10){\circle*{1}}
 \put (120,20){\circle*{1}}
 \put (120,30){\circle*{1}}
 \put (120,40){\circle*{1}}    
 \put (120,50){\circle*{1}}
 \put (120,60){\circle*{1}}  
 \put (120,70){\circle*{1}}
 \put (120,80){\circle*{1}}    
 \put (120,90){\circle*{1}}
 \put (120,100){\circle*{1}}    
 \put (120,110){\circle*{1}} 
\end{picture}

%% file: 13-24.latex
%
%
%
%
%
\setlength{\unitlength}{3947sp}
\begin{picture}(160,120)(-20,10)
 \put (0,0){\circle*{25}}
 \put (120,0){\circle*{25}}
 \put (0,120){\circle*{25}}
 \put (120,120){\circle*{25}}

 \put (10,0){\circle*{1}}
 \put (20,0){\circle*{1}}
 \put (30,0){\circle*{1}}
 \put (40,0){\circle*{1}}    
 \put (50,0){\circle*{1}}
 \put (60,0){\circle*{1}}  
 \put (70,0){\circle*{1}}
 \put (80,0){\circle*{1}}    
 \put (90,0){\circle*{1}}
 \put (100,0){\circle*{1}}    
 \put (110,0){\circle*{1}}
 
 \put (10,120){\circle*{1}}
 \put (20,120){\circle*{1}}
 \put (30,120){\circle*{1}}
 \put (40,120){\circle*{1}}    
 \put (50,120){\circle*{1}}
 \put (60,120){\circle*{1}}  
 \put (70,120){\circle*{1}}
 \put (80,120){\circle*{1}}    
 \put (90,120){\circle*{1}}
 \put (100,120){\circle*{1}}    
 \put (110,120){\circle*{1}} 
    
\end{picture}

%% file: 14-23_julia.latex
%
%
\setlength{\unitlength}{3947sp}
\begin{picture}(160,120)(-20,10)
 \put (0,0){\circle*{25}}
 \put (120,0){\circle*{25}}
 \put (0,120){\circle*{25}}
 \put (120,120){\circle*{25}}

 

 \put (110,10){\circle*{1}}
 \put (100,20){\circle*{1}}
 \put (90,30){\circle*{1}}
 \put (80,40){\circle*{1}}
 \put (70,50){\circle*{1}}
 \put (60,60){\circle*{1}}    
 \put (50,70){\circle*{1}}
 \put (40,80){\circle*{1}}  
 \put (30,90){\circle*{1}}
 \put (20,100){\circle*{1}}    
 \put (10,110){\circle*{1}}

 \put (10,10){\circle*{1}}
 \put (20,20){\circle*{1}}
 \put (30,30){\circle*{1}}
 \put (40,40){\circle*{1}}    
 \put (50,50){\circle*{1}}
 \put (60,60){\circle*{1}}  
 \put (70,70){\circle*{1}}
 \put (80,80){\circle*{1}}    
 \put (90,90){\circle*{1}}
 \put (100,100){\circle*{1}}
 \put (110,110){\circle*{1}}    
\end{picture}

%% file: rect.latex
%
%
\setlength{\unitlength}{2750sp}
\begin{picture}(89,160)(-5,20)
 \put (0,0){\line(1,0){80}}
 \put (0,80){\line(1,0){80}}
 \put (0,160){\line(1,0){80}}
  
 \put (0,0){\line(0,1){160}}
 \put (80,0){\line(0,1){160}}
\end{picture}

%% file: 13-45.latex
%
%
\setlength{\unitlength}{3947sp}
\begin{picture}(130,160)(-15,60)
 \put (0,0){\circle*{25}}
 \put (100,0){\circle*{25}}
 \put (0,100){\circle*{25}}
 \put (100,100){\circle*{25}}
 \put (0,200){\circle*{25}}
 \put (100,200){\circle*{25}}
 

 \put (0,10){\circle*{1}}
 \put (0,20){\circle*{1}}
 \put (0,30){\circle*{1}}
 \put (0,40){\circle*{1}}    
 \put (0,50){\circle*{1}}
 \put (0,60){\circle*{1}}  
 \put (0,70){\circle*{1}}
 \put (0,80){\circle*{1}}    
 \put (0,90){\circle*{1}}

 \put (10,190){\circle*{1}}
 \put (20,180){\circle*{1}}
 \put (30,170){\circle*{1}}
 \put (40,160){\circle*{1}}    
 \put (50,150){\circle*{1}}
 \put (60,140){\circle*{1}}  
 \put (70,130){\circle*{1}}
 \put (80,120){\circle*{1}}    
 \put (90,110){\circle*{1}}

\end{picture}

%% file: 14-35.latex
%
%
\setlength{\unitlength}{3947sp}
\begin{picture}(130,160)(-15,60)
 \put (0,0){\circle*{25}}
 \put (100,0){\circle*{25}}
 \put (0,100){\circle*{25}}
 \put (100,100){\circle*{25}}
 \put (0,200){\circle*{25}}
 \put (100,200){\circle*{25}}
 

\put (0,110){\circle*{1}}
 \put (0,120){\circle*{1}}
 \put (0,130){\circle*{1}}
 \put (0,140){\circle*{1}}    
 \put (0,150){\circle*{1}}
 \put (0,160){\circle*{1}}  
 \put (0,170){\circle*{1}}
 \put (0,180){\circle*{1}}    
 \put (0,190){\circle*{1}}

 \put (10,10){\circle*{1}}
 \put (20,20){\circle*{1}}
 \put (30,30){\circle*{1}}
 \put (40,40){\circle*{1}}    
 \put (50,50){\circle*{1}}
 \put (60,60){\circle*{1}}  
 \put (70,70){\circle*{1}}
 \put (80,80){\circle*{1}}    
 \put (90,90){\circle*{1}}
    
\end{picture}

%% file: 13-46.latex
%
%
%
%
%
\setlength{\unitlength}{3947sp}
\begin{picture}(130,160)(-15,60)
 \put (0,0){\circle*{25}}
 \put (100,0){\circle*{25}}
 \put (0,100){\circle*{25}}
 \put (100,100){\circle*{25}}
 \put (0,200){\circle*{25}}
 \put (100,200){\circle*{25}}
 
 
 \put (0,10){\circle*{1}}
 \put (0,20){\circle*{1}}
 \put (0,30){\circle*{1}}
 \put (0,40){\circle*{1}}    
 \put (0,50){\circle*{1}}
 \put (0,60){\circle*{1}}  
 \put (0,70){\circle*{1}}
 \put (0,80){\circle*{1}}    
 \put (0,90){\circle*{1}}
 
 \put (100,110){\circle*{1}}
 \put (100,120){\circle*{1}}
 \put (100,130){\circle*{1}}
 \put (100,140){\circle*{1}}    
 \put (100,150){\circle*{1}}
 \put (100,160){\circle*{1}}  
 \put (100,170){\circle*{1}}
 \put (100,180){\circle*{1}}    
 \put (100,190){\circle*{1}}

\end{picture}

%% file: 14-36.latex
%
%
\setlength{\unitlength}{3947sp}
\begin{picture}(130,160)(-15,60)
 \put (0,0){\circle*{25}}
 \put (100,0){\circle*{25}}
 \put (0,100){\circle*{25}}
 \put (100,100){\circle*{25}}
 \put (0,200){\circle*{25}}
 \put (100,200){\circle*{25}}
 
 \put (10,10){\circle*{1}}
 \put (20,20){\circle*{1}}
 \put (30,30){\circle*{1}}
 \put (40,40){\circle*{1}}    
 \put (50,50){\circle*{1}}
 \put (60,60){\circle*{1}}  
 \put (70,70){\circle*{1}}
 \put (80,80){\circle*{1}}    
 \put (90,90){\circle*{1}}
 \put (95,95){\circle*{1}}

 \put (10,110){\circle*{1}}
 \put (20,120){\circle*{1}}
 \put (30,130){\circle*{1}}
 \put (40,140){\circle*{1}}    
 \put (50,150){\circle*{1}}
 \put (60,160){\circle*{1}}  
 \put (70,170){\circle*{1}}
 \put (80,180){\circle*{1}}    
 \put (90,190){\circle*{1}}
 \put (95,195){\circle*{1}}
 
\end{picture}

%% file: rectt.latex
%
%
\setlength{\unitlength}{2750sp}
\begin{picture}(169,160)(-5,20)
 \put (0,0){\line(1,0){80}}
 \put (0,80){\line(1,0){80}}
 \put (0,160){\line(1,0){80}}
  
 \put (0,0){\line(0,1){160}}
 \put (80,0){\line(0,1){160}}

 \put (80,0){\line(1,0){80}}
 \put (80,80){\line(1,0){80}}
 \put (80,160){\line(1,0){80}}

 \put (160,0){\line(0,1){160}}
\end{picture}

%% file: 14-36_julia.latex
%
%
\setlength{\unitlength}{3947sp}
\begin{picture}(240,160)(-15,60)
 \put (0,0){\circle*{25}}
 \put (100,0){\circle*{25}}
 \put (200,0){\circle*{25}}
 \put (0,100){\circle*{25}}
 \put (100,100){\circle*{25}}
 \put (200,100){\circle*{25}}
 \put (0,200){\circle*{25}}
 \put (100,200){\circle*{25}}
 \put (200,200){\circle*{25}}
 
 \put (10,10){\circle*{1}}
 \put (20,20){\circle*{1}}
 \put (30,30){\circle*{1}}
 \put (40,40){\circle*{1}}    
 \put (50,50){\circle*{1}}
 \put (60,60){\circle*{1}}  
 \put (70,70){\circle*{1}}
 \put (80,80){\circle*{1}}    
 \put (90,90){\circle*{1}}
 \put (95,95){\circle*{1}}

 \put (10,110){\circle*{1}}
 \put (20,120){\circle*{1}}
 \put (30,130){\circle*{1}}
 \put (40,140){\circle*{1}}    
 \put (50,150){\circle*{1}}
 \put (60,160){\circle*{1}}  
 \put (70,170){\circle*{1}}
 \put (80,180){\circle*{1}}    
 \put (90,190){\circle*{1}}
 \put (95,195){\circle*{1}}

 \put (110,10){\circle*{1}}
 \put (120,20){\circle*{1}}
 \put (130,30){\circle*{1}}
 \put (140,40){\circle*{1}}
 \put (150,50){\circle*{1}}
 \put (160,60){\circle*{1}}
 \put (170,70){\circle*{1}}
 \put (180,80){\circle*{1}}
 \put (190,90){\circle*{1}}
 \put (195,95){\circle*{1}} 
\end{picture}

%% file: 14-36_julia3.latex
%
%
\setlength{\unitlength}{3947sp}
\begin{picture}(240,160)(-15,60)
 \put (0,0){\circle*{25}}
 \put (100,0){\circle*{25}}
 \put (200,0){\circle*{25}}
 \put (0,100){\circle*{25}}
 \put (100,100){\circle*{25}}
 \put (200,100){\circle*{25}}
 \put (0,200){\circle*{25}}
 \put (100,200){\circle*{25}}
 \put (200,200){\circle*{25}}
 
 \put (10,10){\circle*{1}}
 \put (20,20){\circle*{1}}
 \put (30,30){\circle*{1}}
 \put (40,40){\circle*{1}}    
 \put (50,50){\circle*{1}}
 \put (60,60){\circle*{1}}  
 \put (70,70){\circle*{1}}
 \put (80,80){\circle*{1}}    
 \put (90,90){\circle*{1}}
 \put (95,95){\circle*{1}}

 \put (190,110){\circle*{1}}
 \put (180,120){\circle*{1}}
 \put (170,130){\circle*{1}}    
 \put (160,140){\circle*{1}}
 \put (150,150){\circle*{1}}  
 \put (140,160){\circle*{1}}
 \put (130,170){\circle*{1}}    
 \put (120,180){\circle*{1}}
 \put (110,190){\circle*{1}}

 \put (90,10){\circle*{1}}
 \put (80,20){\circle*{1}}
 \put (70,30){\circle*{1}}
 \put (60,40){\circle*{1}}
 \put (50,50){\circle*{1}}
 \put (40,60){\circle*{1}}
 \put (30,70){\circle*{1}}
 \put (20,80){\circle*{1}}
 \put (10,90){\circle*{1}}
\end{picture}

%% file: 14-23.latex
%
%
\setlength{\unitlength}{3947sp}
\begin{picture}(160,120)(-20,10)
 \put (0,0){\circle*{25}}
 \put (120,0){\circle*{25}}
 \put (0,120){\circle*{25}}
 \put (120,120){\circle*{25}}
 
 \put (10,10){\circle*{1}}
 \put (20,20){\circle*{1}}
 \put (30,30){\circle*{1}}
 \put (40,40){\circle*{1}}    
 \put (50,50){\circle*{1}}
 \put (60,60){\circle*{1}}  
 \put (70,70){\circle*{1}}
 \put (80,80){\circle*{1}}    
 \put (90,90){\circle*{1}}
 \put (100,100){\circle*{1}}    
 \put (110,110){\circle*{1}}
 
 \put (10,110){\circle*{1}}
 \put (20,100){\circle*{1}}
 \put (30,90){\circle*{1}}
 \put (40,80){\circle*{1}}    
 \put (50,70){\circle*{1}}
 \put (60,60){\circle*{1}}  
 \put (70,50){\circle*{1}}
 \put (80,40){\circle*{1}}    
 \put (90,30){\circle*{1}}
 \put (100,20){\circle*{1}}    
 \put (110,10){\circle*{1}} 
\end{picture}